\documentclass[12pt]{article}
\usepackage{amsfonts}
\usepackage{amsmath}
\usepackage{graphicx}
\usepackage[english]{babel}
\usepackage[]{graphicx}
\usepackage{color}
\usepackage{url}
\usepackage{html}
\usepackage{bbm}

\title{Particle Kinematics in Ho\v{r}ava-Lifshitz Gravity}
\author{D.~Capasso\footnote{dcapass00@ccny.cuny.edu},  A.~P.~Polychronakos\footnote{alexios@sci.ccny.cuny.edu}\\
Physics Department, City College of the CUNY\\
160 Convent Avenue, New York, NY 10031}
\begin{document}

\maketitle
\hfill\raisebox{200pt}[0pt][0pt]{CCNY-HEP-09/04}
\begin{abstract}
We study the deformed kinematics of point particles in the Ho\v{r}ava theory of gravity. This is achieved by considering particles as the optical limit of fields with a generalized Klein-Gordon action. We derive the deformed geodesic equation and study in detail the cases of flat and spherically symmetric (Schwarzschild-like) spacetimes.
As the theory is not invariant under local Lorenz transformations, deviations from standard kinematics become evident even for flat manifolds, supporting superluminal as well as massive luminal particles. These deviations from standard behavior could be used for experimental tests of this modified theory of gravity.
%We study a deformed kinematics that arises as optical of a deformed Klein-Gordon theory which was used to introduce generalized matter fields in the Ho\v{r}ava-Lifshitz theory of gravity. The new matter field action is not invariant anymore under the full Poicar\'e group and this will become evident in studying the kinematics on a flat Minkowskian manifold. All the modification to the usual kinematics highlighted here are small for low energy particles, recovering the Poincar\'e symmetry in the IR limit. Although this small deviations may be difficult to observe, the theory makes possible, in principle, the existence of luminal massive particles and superluminal particles.
\end{abstract}

\tableofcontents

\section{Introduction}\label{sec:intro}

One of the main problems in the quantization of the Hilbert-Einstein theory of Gravity is that it yields a non-renormalizable quantum field theory. To solve this problem theories with higher derivatives and higher powers of the curvature were considered; the Ho\v{r}ava-Lifshitz theory introduced in \cite{Horava:MQC,Horava:QGLP} is one such theory.

The Ho\v{r}ava gravity is constructed as an UV completion of the Hilbert-Einstein Gravity which treats space and time differently. The basic property of such a theory is the invariance under the anisotropic rescaling
\[
x\to bx \qquad t\to b^{z}t,
\]
which makes the conformal dimensions ($[\phantom{-}]_{s}$) of space and time to be different:
\[
[x]_{s}=-1 \qquad [t]_{s}=-z.
\]
The reason in doing this is that, choosing an action invariant under the deformed rescaling with an appropriate value of $z$, the action will turn out to be power-counting renormalizable.

Because of the anisotropy in the rescaling, space-time must be of the form $M=\mathbb{R}\times\Sigma$ where $\Sigma$ is a space-like and, for simplicity, compact $3$-dimensional surface. To ensure this property the manifold $M$ possesses a codimension-one foliation\footnote{A codimension-q foliation $\mathcal{F}$ on a $d$-dimensional manifold $M$ means that there exists an atlas $(y^{a},x^{i})$ with $a=1,...,q$ and $i=1,...,d-q$ such that the transition function acts as follows
\[
x^{i}\to \tilde{x}^{i}=\tilde{x}^{i}(x,y) \qquad
y^{a}\to \tilde{y}^{a}=\tilde{y}^{a}(y),
\]
that is, we consider the action only of diffeomorphisms that leaves unchanged the foliation structure (\cite{Lawson:F,MoerdijkMrcun:IFLG}).} structure $\mathcal{F}$. Therefore, the Ho\v{r}ava-Lifshitz theory was  constructed as a theory invariant only under the diffeomorfisms that leave the foliation structure unchanged, that is,
\[
x^{i}\to \tilde{x}^{i}=\tilde{x}^{i}(x,t) \qquad
t\to \tilde{t}=\tilde{t}(t).
\]

The space-time metric $g_{\mu\nu}$, because of the foliation structure, can be globally decomposed in terms of the ADM decomposition:
\begin{equation}
g_{\mu\nu}=
\left(\begin{array}{cc}
-N^{2}+N_{i}N^{i} & N_{j}\\
N_{i} & h_{ij}
\end{array}\right) \quad
g^{\mu\nu}=
\left(\begin{array}{cc}
-\frac{1}{N^{2}} & \frac{N^{j}}{N^{2}}\\
\frac{N^{i}}{N^{2}} & h^{ij}-\frac{N^{i}N^{j}}{N^{2}}
\end{array}\right)
\end{equation}
where $h_{ij}(x,t)$ is the metric on $\Sigma$ and $N(x,t)$ and $N_{i}(x,t)$ are called, respectively, lapse and shift functions. The general structure is given by the metric $g^{\mu\nu}$ and a time-like vector $n^{\alpha}$ orthogonal to the space-like hypersurface $\Sigma$. The metric is related to $n^{\alpha}$ through the relation \cite{Wald:GR}
\[
g^{\alpha\beta}=h^{\alpha\beta}-n^{\alpha}n^{\beta}
\]
where $h^{\alpha\beta}$ corresponds to the embedded metric of the space-like surface $\Sigma$. The lapse and shift functions are defined by the relation
\[
Nn^{\alpha}=t^{\alpha}-N^{\alpha}
\]
with $N^{\alpha}$ tangent to the space-like surface ($N^{\alpha}n_{\alpha}=0$), while $t^{\alpha}$ is the time-like vector tangent to a continuous set of geodesics that individuate globally the time direction, allowing the foliation. Moreover $h_{\alpha}^{\phantom{-}\beta}$ and $-n_{\alpha}n^{\beta}$ are, respectively, the projector on $\Sigma$ and its orthogonal projector. The ADM decomposition introduced above corresponds to the particular choice of coordinates in which $t^{\alpha}\equiv\frac{\partial}{\partial t}$. In particular, in the ADM decomposition, the projector on $\Sigma$ takes the form
\begin{equation}\label{ADMh_projector}
h_{\alpha}^{\phantom{-}\beta}=\left(\begin{array}{cc}
0 & N^{j}\\
0 & \delta_{i}^{\phantom{-}j}
\end{array}\right).
\end{equation}

In principle such a metric makes sense only if there exists a space-time structure. In the Ho\v{r}ava-Lifshitz model the usual General Relativity is recovered in the IR limit and hence also the space-time must be seen as emerging from the theory in the low energy limit. In general, and in particular in the UV limit, the only quantity that can be interpreted as a metric is $h_{ij}$, that is, the embedding of $h_{\alpha\beta}$ on $\Sigma$. The matrix $g_{\alpha\beta}$ cannot be in general interpreted as a metric in the usual sense because a free particle may not, in general, move along geodesics determined by $g_{\alpha\beta}$. This happens because in this theory the time-like direction $n_{\alpha}$, which allows a distinction between space and time, is also considered as a degree of freedom.

The Ho\v{r}ava-Lifshitz action
\begin{equation}\label{HLaction}
S_{HL}=S_{K}-S_{V}
\end{equation}
contains a kinetic term $S_{K}$, involving time derivatives, and a potential term $S_{V}$, involving only space derivatives. In the potential term there are also higher space derivatives, as well as higher powers of the $3$-dimensional curvature on $\Sigma$. The full Ho\v{r}ava-Lifshitz action, in order of descending dimensions, is
\begin{equation}
\begin{array}{c}
S=\int dtd^{3}x\sqrt{h}N\left\{\frac{2}{\kappa^{2}}(K_{ij}K^{ij}-\lambda K^{2})
-\frac{\kappa^{2}}{2\omega^{4}}C_{ij}C^{ij}
+\frac{\kappa^{2}\mu}{2\omega^{2}}\varepsilon^{ijk}\mathcal{R}_{il}\nabla_{j}\mathcal{R}^{l}_{\phantom{-}k}\right.
\\
\left.-\frac{\kappa^{2}\mu^{2}}{8}\mathcal{R}_{ij}\mathcal{R}^{ij}
+\frac{\kappa^{2}\mu^{2}}{8(1-3\lambda)}\left(\frac{1-4\lambda}{4}\mathcal{R}^{2}
+\Lambda_{W}\mathcal{R}-3\Lambda_{W}^{2}\right)\right\}
\end{array}
\end{equation}
where the kinetic term corresponds to the first bracket, in which
\[
K_{ij}=\frac{1}{2N}(\dot{h}_{ij}-\nabla_{i}N_{j}-\nabla_{j}N_{i}),
\]
$\mathcal{R}_{ij}$ are the spatial components of the Ricci tensor on $\Sigma$, $\mathcal{R}$ is its trace and $C_{ij}$ are the spatial components of the Cotton tensor. The potential term was first introduced using the detailed balance condition, that is, it corresponds to the variation of an action describing a $3$-dimensional Euclidean gravity \cite{Horava:QGLP}.

Such a theory has a UV critical point $z=3$ and an IR critical point $z=1$ that corresponds to the relativistic case; indeed, in the IR limit $w\to\infty$ and the quadratic terms in the curvature go to zero obtaining
\[
S=\int dtd^{3}x\sqrt{h}N\left\{\frac{2}{\kappa^{2}}(K_{ij}K^{ij}-\lambda K^{2})
+\frac{\kappa^{2}\mu^{2}}{8(1-3\lambda)}\Lambda_{W}\left(\mathcal{R}-3\Lambda_{W}\right)\right\},
\]
which is isotropic under the rescaling of space ant time. Comparing the IR limit of the Ho\v{r}ava-Lifshitz action to the Einstein-Hilbert action
\begin{equation}
S_{EH}=\frac{1}{16\pi G}\int\sqrt{g}d^{4}x[\mathcal{R}-2\Lambda_{E}]=
\frac{1}{16\pi G}\int\sqrt{h}d^{4}xN[K_{ij}K^{ij}-K^{2}+\mathcal{R}-2\Lambda_{E}]
\end{equation}
we obtain, respectively, the emergent velocity of light, the emergent Newton constant and the cosmological constant
\[
c=\frac{\kappa^{2}\mu}{4}\sqrt{\frac{\Lambda_{W}}{1-3\lambda}} \quad ([c]_{s}=2(z-1)),
\qquad
G_{N}=\frac{\kappa^{2}}{32\pi c} \quad ([G_{N}]_{s}=-2),
\qquad
\Lambda=\frac{3}{2}\Lambda_{W}.
\]
The IR limit of the Ho\v{r}ava-Lifshitz action will recover Einstein Gravity only if the running constant $\lambda$ becomes $1$ in the $z=1$ fixed point.

The detailed balance condition is a way to construct interacting terms in the Lifshitz scalar theory, but terms that break softly the detailed balance condition can be also considered. With softly breaking we mean that the UV limit is still described by the potential obtained from the detailed balance condition.

The equations of motion of the action (\ref{HLaction}) were obtained in \cite{LuMeiPope:SHG,KiritsisKofinas:HLC}, while several aspects relative to the spherically symmetric solution \cite{MyungKim:THLBH,CaiCaoOhta:TBHHLG,CaiCaoOhta:ThBHHLG,GhodsiHatefi:ERSHG,ColgainYavartanoo:DSHLG} and to cosmology \cite{Calcagni:CLU,WangWu:TCCMHLTG,Minamitsuji:CCAMHLT,TakahashiSoda:CPGWLP,Saridakis:HLDE,CaiSaridakis:NSCMNRG,LeonSaridakis:PSAHLC,CarloniElizaldeSilva:APSHLC}, as well as other fundamental aspects \cite{LiPang:THLG,BogdanosSaridakis:PIHG,Calcagni:DBHLG,OrlandoReffert:RHLG} of this theory, were analyzed.

In this context a matter field was introduced in \cite{ChenHuang:FTLP} mimicking the Lifshitz scalar field theory (for an introduction on the Lifshitz scalar field theory see \cite{Horava:MQC,Horava:QGLP}). The proposed scalar field theory was
\begin{equation}\label{SMatter}
S_{M}=\frac{1}{2}\int d^{4}x\sqrt{h}N\left\{
\frac{1}{N^{2}}(\partial_{t}\phi-N^{i}\partial_{i}\phi)^{2}-\sum_{J\geq2}\mathcal{O}_{J}\star\phi^{J}
\right\}
\end{equation}
where we distinguish between a kinetic term, contained in the first bracket, and a potential term involving only spatial derivatives, where
\[
O_{J}=\sum_{n=0}^{n_{J}}(-1)^{n}\frac{\lambda_{J,n}}{M^{2n+\frac{d-1}{2}J-d-1}}\Delta^{n}
\]
and the $\star$ product represents all the possible combinations in the application of $\Delta=h^{ij}D_{i}D_{j}$, being $D_{j}$ the covariant derivative\footnote{The derivative $D_{\alpha}$ is the projection of the covariant derivative on the space-like surface
\[
D_{\alpha}T_{\mu_{1}...\mu_{m}}^{\nu_{1}...\nu_{n}}=
h^{\phantom{-}\beta}_{\alpha}h^{\phantom{-}\lambda_{1}}_{\mu_{1}}...h^{\phantom{-}\lambda_{m}}_{\mu_{m}}
h^{\phantom{-}\nu_{1}}_{\rho_{1}}...h^{\phantom{-}\nu_{n}}_{\rho_{n}}\nabla_{\beta}T_{\lambda_{1}...\lambda_{m}}^{\rho_{1}...\rho_{n}}
\]
where $h^{\phantom{-}\alpha}_{\beta}$ is the projector onto the space-like hypersurface $\Sigma$. Using the orthogonality of $n_{\alpha}$ respect to the space-like hypersurface $\Sigma$, that is, $h^{\phantom{-}\alpha}_{\beta}n_{\alpha}=0$, it is simple to show $D_{\alpha}h^{\mu\nu}=0$.} on $\Sigma$, to the $\phi$'s, i.e.
\[
\Delta^{2}\star\phi^{3}=c_{1}(\Delta\phi)^{2}\phi+c_{2}\phi^{2}\Delta^{2}\phi
\]
with $c_{1},c_{2}$ constants. In \cite{ChenHuang:FTLP} all the conditions necessary to have a power-counting renormalizable theory are derived. Here we will be interested only in the effective mass term corresponding to $J=2$, to describe just the motion of matter without any interaction:
\begin{equation}\label{SMatterJ=2}
S_{M}=\frac{1}{2}\int d^{4}x\sqrt{h}N\left\{
\frac{1}{N^{2}}(\partial_{t}\phi-N^{i}\partial_{i}\phi)^{2}
-\sum_{n=0}^{z}(-1)^{n}\frac{\lambda_{2,n}}{M^{2(n-1)}}\Delta^{n}\star\phi^{2}
\right\}.
\end{equation}

Henceforth we will consider the modified Ho\v{r}ava-Lifshitz theory described by the Kehagias-Sfetsos action
\begin{eqnarray}\label{KSaction}
S &=&\int dtd^{3}x\sqrt{h}N\left\{\frac{2}{\kappa^{2}}(K_{ij}K^{ij}-\lambda K^{2})
-\frac{\kappa^{2}}{2\omega^{4}}C_{ij}C^{ij}
+\frac{\kappa^{2}\mu}{2\omega^{2}}\epsilon^{ijk}\mathcal{R}_{il}\nabla_{j}\mathcal{R}^{l}_{\phantom{-}k}+\right.\nonumber\\
&& \left.-\frac{\kappa^{2}\mu^{2}}{8}\mathcal{R}_{ij}\mathcal{R}^{ij}
+\frac{\kappa^{2}\mu^{2}}{8(1-3\lambda)}\left(\frac{1-4\lambda}{4}\mathcal{R}^{2}+\Lambda_{W}\mathcal{R}-3\Lambda_{W}^{2}\right)
+\mu^{4}\mathcal{R}\right\}
\end{eqnarray}
introduced in \cite{KehagiasSfetsos:BHFRWGNRG}. This action contains, in addition to the original Ho\v{r}ava action, a soft violation of the detailed balance condition (the term $\mu^{4}\mathcal{R}$). This new term, as observed in \cite{KehagiasSfetsos:BHFRWGNRG}, makes the action have a well behaved limit
\[
\Lambda_{W}\to0
\]
and admits a Minkowski vacuum solution.

As for the original Ho\v{r}ava action, several aspects of the Kehagias-Sfetsos action were analyzed: black holes \cite{Myung:TBHDHLG,Myung:EBHDHLG,LeeKimMyung:EBHHLG,PengWu:HRBHIMHLG}, cosmological solutions \cite{Park:BHSIRMHG}, possible tests \cite{Konoplya:TCHLG,HarkoKivacsLobo:SSTHLG,Park:THGDE,IorioRuggiero:HLGSSOM} and fundamental aspects of the theory \cite{CharmousisNizPadillaSaffin:SCHG,BlasPujolasSibiryakov:EMIHG}.

In \cite{GermaniKehagiasSfetsos:RQGLP} the authors rewrite the action (\ref{KSaction}) in a covariant way considering $g^{\alpha\beta}$ and the time-like covector $n_{\alpha}$ as the fields of the theory instead of the ADM components of the metric. The covector $n_{\alpha}$ is such that
\[
n_{\alpha}=-N\partial_{\alpha}\chi \qquad n_{\alpha}n^{\alpha}=-1
\]
where the lapse function $N$ encodes the normalization and $\chi$ parametrizes the foliation, that is, $\chi$ has a different constant value on each foliation. The defining condition $n_{\alpha}=-N\partial_{\alpha}\chi$ verifies the Frobenius integrability condition
\[
\mathcal{F}_{\mu\nu}=D_{[\mu}n_{\nu]}=0,
\]
which means that a zero vorticity condition for $n_{\alpha}$ is satisfied; this a necessary condition to have a foliation structure.

In the original Ho\v{r}ava action, $K_{ij}$ were the spatial components of the induced extrinsic curvature
\[
K_{\alpha\beta}=\frac{1}{2}\mathcal{L}_{n}h_{\alpha\beta}.
\]
Noting that $K_{\alpha\beta}n^{\beta}=\frac{1}{2}\mathcal{L}_{n}(h_{\alpha\beta}n^{\beta})=0$, we simply deduce that $K_{\alpha\beta}=h_{\alpha}^{\phantom{-}\gamma}h_{\beta}^{\phantom{-}\delta}K_{\gamma\delta}$, which reduces to
\[
K_{\alpha\beta}=h_{\alpha}^{\phantom{-}i}h_{\beta}^{\phantom{-}j}K_{ij},
\]
$h_{\alpha}^{\phantom{-}\beta}$ being zero for $\beta$ corresponding to the time component. Therefore, in the ADM decomposition, we have
\[
K_{\alpha\beta}K^{\alpha\beta}=K_{ij}K^{ij} \quad\textrm{and}\quad
K=K_{\alpha\beta}g^{\alpha\beta}=K_{ij}h^{ij};
\]
then
\[
K_{\alpha\beta}K^{\alpha\beta}-\lambda K^{2}=
K_{ij}K^{ij}-\lambda(K^{i}_{\phantom{-}i})^{2}
\]
reproduces exactly the kinetic term in the Ho\v{r}ava-Lifshitz action. Moreover, noting that the $3$-dimensional curvature can be written as
\[
\mathcal{R}_{\lambda\mu\nu\rho}=R_{\alpha\beta\gamma\delta}h^{\phantom{-}\alpha}_{\lambda}h^{\phantom{-}\beta}_{\mu}h^{\phantom{-}\gamma}_{\nu}h^{\phantom{-}\delta}_{\rho}-
2K_{\mu[\nu}K_{\rho]\lambda}
\]
and the Cotton tensor is defined as
\[
C^{\mu\nu}=\eta^{\mu\alpha\beta}D_{\alpha}\left[\mathcal{R}^{\nu}_{\phantom{-}\beta}-\frac{1}{4}\mathcal{R}\delta^{\nu}_{\phantom{-}\beta}\right]
\]
where $\eta^{\mu\alpha\beta}\equiv\eta^{\mu\alpha\beta\delta}n_{\delta}$ is the $3$-dimensional volume form, it is simple to show that
\[
\mathcal{R}_{\alpha\beta}=h_{\alpha}^{\phantom{-}i}h_{\beta}^{\phantom{-}j}\mathcal{R}_{ij}
\quad
\mathcal{R}=\mathcal{R}_{\alpha\beta}g^{\alpha\beta}=\mathcal{R}_{ij}h^{ij}
\quad
C_{\alpha\beta}=h_{\alpha}^{\phantom{-}i}h_{\beta}^{\phantom{-}j}C_{ij}.
\]
As for the kinetic term we can promote the $3$-dimensional indices in the action (\ref{KSaction}) to $4$-dimensional indices obtaining the same expression in the case in which the theory is written in the ADM components. Hence the authors of \cite{GermaniKehagiasSfetsos:RQGLP} generalize the action (\ref{KSaction}) in a diffeomorphism-invariant form
\begin{eqnarray}\label{RKSaction}
S_{GKS} &=& \int d^{4}x\sqrt{-g}\left\{
\frac{2}{\kappa^{2}}(K_{\alpha\beta}K^{\alpha\beta}-\lambda K^{2})
-\frac{\kappa^{2}}{2\omega^{4}}C_{\alpha\beta}C^{\alpha\beta}
+\frac{\kappa^{2}\mu}{2\omega^{2}}C_{\alpha\beta}\mathcal{R}^{\alpha\beta}
-\frac{\kappa^{2}\mu^{2}}{8}\mathcal{R}_{\alpha\beta}\mathcal{R}^{\alpha\beta}\right.\nonumber\\
&& \left.+\frac{\kappa^{2}\mu^{2}}{8(1-3\lambda)}\left(\frac{1-4\lambda}{4}\mathcal{R}^{2}+\Lambda_{W}\mathcal{R}-3\Lambda_{W}\right)
+\eta^{4}\mathcal{R}+\mathcal{L}_{norm}\right\}
\end{eqnarray}
where the Frobenius integrability condition and the normalization of $n_{\alpha}$ are ensured by
\[
\mathcal{L}_{norm}=
B^{\alpha\beta}\mathcal{F}_{\alpha\beta}
+M^{\alpha\beta\mu\nu}B_{\alpha\beta}B_{\mu\nu}
+\rho(N^{2}\partial_{\alpha}\chi\partial^{\alpha}\chi+1)
\]
where $M^{\alpha\beta\mu\nu}$, $B^{\alpha\beta}$ and $\rho$ are Lagrange multipliers.

Here we will be interested in the case of zero cosmological constant and $\lambda=1$. 
In this case the theory has a Minkowski vacuum solution as is evident from the spherical solution in \cite{KehagiasSfetsos:BHFRWGNRG} in the case we set $M=0$.

To study the transformation properties of a matter field we generalize the action (\ref{SMatterJ=2}) to the following diffeomorphism-invariant action
\begin{eqnarray}\label{RSMatterJ=2}
S_{M} &=& \frac{1}{2}\int d^{4}x\sqrt{-g}\left\{
-g^{\mu\nu}\partial_{\mu}\phi\partial_{\nu}\phi^{*}
-\lambda_{2,0}M^{2}\phi\phi^{*}
+\left(\lambda_{2,1}-\frac{1}{2}\right)(\phi\Delta\phi^{*}+\phi^{*}\Delta\phi)\right.\nonumber\\
&&\left.-\sum_{n=2}^{z}\sum_{k=0}^{n}(-1)^{n}\frac{\lambda_{2,n,k}}{M^{2(n-1)}}\Delta^{n-k}\phi\Delta^{k}\phi^{*}
\right\}
\end{eqnarray}
where the Laplace operator is redefined as
\[
\Delta\equiv h^{\alpha\beta}D_{\alpha}D_{\beta} \quad
(=h^{\alpha\beta}\nabla_{\alpha}\nabla_{\beta})
\]
and $\lambda_{2,n,k}=\lambda_{2,n,n-k}$ to make the action real. The action (\ref{RSMatterJ=2}) it is a generalization of the Klein-Gordon action where the terms involving the various $\lambda$'s (excluding the term $\lambda_{2,0}M^{2}\phi\phi^{*}$ that corresponds to the usual mass term in the Klein-Gordon equation) must be considered as small corrections.

Note that the action (\ref{RSMatterJ=2}) reduces to the original action (\ref{SMatterJ=2}) for a Minkowski space-time in ADM coordinates, which corresponds to the case studied in \cite{ChenHuang:FTLP}.

In section~\ref{sec:optical_limit} we will find the optical limit approximation of the scalar theory (\ref{RSMatterJ=2}) obtaining the general equation of motion of massive and massless particles. In particular we will study the kinematics in a flat space-time in section~\ref{sec:Minkowski} and the dynamics in a static spherical symmetric space-time in section~\ref{sec:Spherical}.

\section{The Optical Limit}\label{sec:optical_limit}
To obtain the ray optical structure\footnote{The ray optical structure $H$ is the Hamiltonian of our system. Moreover the disperison relation will be given by the condition $H=0$. (See \cite{Perlick:ROFPAGR} for a review.)} which describes the optical limit behavior, we write the equation of motion for the scalar field $\phi$ obtained from the action (\ref{RSMatterJ=2}) and express the scalar field as
\begin{equation}\label{plane_wave}
\phi=Se^{i\psi}
\end{equation}
to find the eikonal equation. To study the ray approximation we will consider that the derivatives of the wavefront $S$ produce a negligible contribution if the curvature is weak enough to consider almost plane wavefronts, and that the four-momentum, defined in terms of the field $\psi$ as $p_{\mu}\equiv\partial_{\mu}\psi$, changes slowly; that is, we will consider higher derivatives of the four-momentum to be negligible. Moreover, because the constants $\tilde{\lambda}_{2,n}$ are small in the IR limit, we can retain only the highest power of the four-momentum in the eikonal equation neglecting all the other terms. For the same reason, and the fact that we are considering metrics that change slowly, we will further make the following approximation:
\begin{equation}\label{approx_metric}
\nabla_{\mu}h^{\mu\nu}=\nabla_{\mu}(n^{\mu}n^{\nu})\simeq0.
\end{equation}
From the eikonal equation we can deduce the ray optical structure $H$ of the theory replacing $\partial_{\mu}\psi$ with the momenta $p_{\mu}$.

Note that in the case of a Minkowski space-time we have exactly
\[
\nabla_{\mu}h^{\mu\nu}=\partial_{\mu}h^{\mu\nu}=0
\]
if we consider only the case $h=\{0,1,1,1\}$ and all other cases obtained by a Lorentz transformation, as we will do in section~\ref{sec:Minkowski}.

The equation of motion obtained from the action (\ref{RSMatterJ=2}), using the results in appendix~\ref{sec:appendix} and the approximation (\ref{approx_metric}), is
\begin{eqnarray}
&g^{\mu\nu}\nabla_{\mu}\nabla_{\nu}\phi
-\lambda_{2,0}M^{2}+\left(\lambda_{2,1}-\frac{1}{2}\right)2\Delta\phi+&\nonumber\\
&-\frac{1}{M^{2}}(2\lambda_{2,2,0}+\lambda_{2,2,1})\Delta^{2}\phi
+\frac{1}{M^{4}}(2\lambda_{2,3,0}+2\lambda_{2,3,1})\Delta^{3}\phi+...=0.&
\end{eqnarray}
Note that the term $\nabla_{\mu}\nabla_{\nu}\phi$, with $\phi$ given by (\ref{plane_wave}), can be expanded as
\[
\nabla_{\mu}\nabla_{\nu}\phi=
\nabla_{\mu}\partial_{\nu}\phi=
\nabla_{\mu}(\partial_{\nu}Se^{i\psi}+i\phi\partial_{\nu}\psi)
\]
and that, considering the approximation that the wave front is locally constant, $\partial_{\nu}S=0$, it becomes
\[
\simeq\nabla_{\mu}(i\phi\partial_{\nu}\psi)=
i(\partial_{\mu}Se^{i\psi}\partial_{\nu}\psi+i\phi\partial_{\mu}\psi\partial_{\nu}\psi+i\phi\partial_{\mu}\partial_{\nu}\psi)\simeq
-\phi\partial_{\mu}\psi\partial_{\nu}\psi
\]
where in the last step we considered that the four-momentum changes slowly. Thus we have the following expansions:
\[
g^{\mu\nu}\nabla_{\mu}\nabla_{\nu}\phi\simeq
-\phi g^{\mu\nu}\partial_{\mu}\psi\partial_{\nu}\psi
\]\[
\Delta\phi=h^{\mu\nu}\nabla_{\mu}\nabla_{\nu}\phi\simeq
-\phi h^{\mu\nu}\partial_{\mu}\psi\partial_{\nu}\psi
\]\[
\Delta^{n}\phi\simeq\Delta^{n-1}(-\phi h^{\mu\nu}\partial_{\mu}\psi\partial_{\nu}\psi)\simeq
-\Delta^{n-1}(\phi)h^{\mu\nu}\partial_{\mu}\psi\partial_{\nu}\psi\simeq
(-1)^{n}(h^{\mu\nu}\partial_{\mu}\psi\partial_{\nu}\psi)^{n}\phi.
\]
Hence the eikonal equation is given by
\begin{equation}
-g^{\mu\nu}\partial_{\mu}\psi\partial_{\nu}\psi-\tilde{\lambda}_{2,0}-\sum_{n=1}^{z}\tilde{\lambda}_{2,n}(h^{\mu\nu}\partial_{\mu}\psi\partial_{\nu}\psi)^{n}=0
\end{equation}
where the new constants $\tilde{\lambda}_{2,n}$\footnote{In the remaining part of the article, when we speak generally of $\tilde{\lambda}_{2,n}$'s or simply of $\tilde{\lambda}$'s we always refer to the $\tilde{\lambda}_{2,n}$'s with $n\geq1$.} are combinations of the old $\lambda$'s appearing in the equation of motion for $\phi$. The constants $\tilde{\lambda}_{2,n}$'s have to be considered small because the relative terms are small corrections to the usual eikonal equation derived from the unmodified Klein-Gordon action. In contrast, $\tilde{\lambda}_{2,0}$ can be simply interpreted as the square of the mass of the particle\footnote{In this context the velocity of light $c$, the emerging velocity of light in the IR limit, is just a conversion constant and will be set to $1$.}. The ray optical structure then is given by the Hamiltonian
\begin{equation}\label{Hmassive}
H=\frac{1}{2\sqrt{\tilde{\lambda}_{2,0}}}\left\{g^{\mu\nu}p_{\mu}p_{\nu}+\tilde{\lambda}_{2,0}+\sum_{n=1}^{z}\tilde{\lambda}_{2,n}(h^{\mu\nu}p_{\mu}p_{\nu})^{n}\right\}=0
\end{equation}
from which we deduce the following equations of motion:
\begin{eqnarray}
\dot{p}_{\alpha}=-\frac{\partial H}{\partial x^{\alpha}} &=&
-\frac{1}{2\sqrt{\tilde{\lambda}_{2,0}}}\left\{
\partial_{\alpha}g^{\mu\nu}p_{\mu}p_{\nu}
+\sum_{n=1}^{z}\tilde{\lambda}_{2,n}n(h^{\rho\lambda}p_{\rho}p_{\lambda})^{n-1}\partial_{\alpha}h^{\mu\nu}p_{\mu}p_{\nu}
\right\}\label{EMpmassive}\\
\dot{x}^{\alpha}=\frac{\partial H}{\partial p_{\alpha}} &=&
\frac{1}{\sqrt{\tilde{\lambda}_{2,0}}}\left\{
g^{\alpha\nu}p_{\nu}
+\sum_{n=1}^{z}\tilde{\lambda}_{2,n}n(h^{\mu\nu}p_{\mu}p_{\nu})^{n-1}h^{\alpha\nu}p_{\nu}
\right\}\label{EMxmassive}
\end{eqnarray}
For the massless case we define the ray optical structure to be
\begin{equation}\label{Hmassless}
H=\frac{1}{2}\left\{g^{\mu\nu}p_{\mu}p_{\nu}+\sum_{n=1}^{z}\tilde{\lambda}_{2,n}(h^{\mu\nu}p_{\mu}p_{\nu})^{n}\right\}=0
\end{equation}
from which we deduce the following equations of motion
\begin{eqnarray}
\dot{p}_{\alpha}=-\frac{\partial H}{\partial x^{\alpha}} &=&
-\frac{1}{2}\left\{\partial_{\alpha}g^{\mu\nu}p_{\mu}p_{\nu}
+\sum_{n=1}^{z}\tilde{\lambda}_{2,n}n(h^{\rho\lambda}p_{\rho}p_{\lambda})^{n-1}\partial_{\alpha}h^{\mu\nu}p_{\mu}p_{\nu}
\right\}\label{EMpmassless}\\
\dot{x}^{\alpha}=\frac{\partial H}{\partial p_{\alpha}} &=&
\left\{
g^{\alpha\nu}p_{\nu}
+\sum_{n=1}^{z}\tilde{\lambda}_{2,n}n(h^{\mu\nu}p_{\mu}p_{\nu})^{n-1}h^{\alpha\nu}p_{\nu}
\right\}\label{EMxmassless}
\end{eqnarray}

In upcoming sections we will analyze the motion of a particle for a Minkowski space-time and for a static spherical symmetric metric. In the next section, instead, we will study the deviations to the geodesic equation due to the deformed kinematics.

\subsection{Corrections to the Geodesic Equation}
The deformed optical structure tells us essentially that the free-falling motion of a particle will only be approximately a geodesic of the metric $g_{\mu\nu}$. Here we will evaluate the first correction in the $\tilde{\lambda}$'s to the geodesic equation. To write down the exact equation of motion we need to invert $\dot{x}^{\alpha}$, finding $p_{\mu}$ as a function of it. Here, because corrections higher than first order in the $\tilde{\lambda}$'s are negligible for low energies, we will retain only terms of first order in the $\tilde{\lambda}$'s.

Differentiating (\ref{EMxmassive}) with respect to the parameter $\tau$ we have:
\[
\ddot{x}^{\alpha}=
\frac{1}{\sqrt{\tilde{\lambda}_{2,0}}}\left[\sqrt{\tilde{\lambda}_{2,0}}g_{\nu\gamma}\dot{x}^{\gamma}\partial_{\beta}g^{\alpha\nu}\dot{x}^{\beta}
-\partial_{\beta}g^{\alpha\nu}\dot{x}^{\beta}\sum_{n=1}^{z}\tilde{\lambda}_{2,n}n(h^{\mu\nu}p_{\mu}p_{\nu})^{n-1}h^{\phantom{-}\gamma}_{\nu}p_{\gamma}\right]
+\frac{1}{\sqrt{\tilde{\lambda}_{2,0}}}\left[g^{\alpha\nu}\dot{p}_{\nu}\right]+
\]\[
+\sum_{n=1}^{z}\frac{\tilde{\lambda}_{2,n}n}{\sqrt{\tilde{\lambda}_{2,0}}}
\left[(n-1)(h^{\mu\nu}p_{\mu}p_{\nu})^{n-2}(\dot{h}^{\mu\nu}p_{\mu}p_{\nu}+2h^{\mu\nu}\dot{p}_{\mu}p_{\nu})h^{\alpha\nu}p_{\nu}
+(h^{\mu\nu}p_{\mu}p_{\nu})^{n-1}(\dot{h}^{\alpha\nu}p_{\nu}
+h^{\alpha\nu}\dot{p}_{\nu})\right]
\]
The momentum $p_{\alpha}$, using the equation of motion (\ref{EMxmassive}), can be expressed as
\[
p_{\alpha}=
\sqrt{\tilde{\lambda}_{2,0}}g_{\alpha\beta}\dot{x}^{\beta}-\sum_{n=1}^{z}\tilde{\lambda}_{2,n}n(h^{\mu\nu}p_{\mu}p_{\nu})^{n-1}h^{\phantom{-}\nu}_{\alpha}p_{\nu}.
\]
Moreover, using the expression above, we have the following approximation:
\[
h^{\mu\nu}p_{\mu}p_{\nu}\simeq
\tilde{\lambda}_{2,0}h_{\mu\nu}\dot{x}^{\mu}\dot{x}^{\nu}+O(\tilde{\lambda}).
\]
Substituting the two expressions above for $p_{\alpha}$ and $h^{\mu\nu}p_{\mu}p_{\nu}$ in (\ref{EMpmassive}) we have
\[
\dot{p}_{\alpha}=
-\frac{1}{2\sqrt{\tilde{\lambda}_{2,0}}}\partial_{\alpha}g^{\mu\nu}p_{\mu}p_{\nu}
-\frac{1}{2\sqrt{\tilde{\lambda}_{2,0}}}\sum_{n=1}^{z}\tilde{\lambda}_{2,n}n(h^{\rho\lambda}p_{\rho}p_{\lambda})^{n-1}\partial_{\alpha}h^{\mu\nu}p_{\mu}p_{\nu}\simeq
\]\[
\frac{\sqrt{\tilde{\lambda}_{2,0}}}{2}[\partial_{\alpha}g_{\beta\delta}\dot{x}^{\beta}\dot{x}^{\delta}]
-\frac{\sqrt{\tilde{\lambda}_{2,0}}}{2}
\left[2\partial_{\alpha}g_{\mu\beta}h^{\mu\gamma}g_{\delta\gamma}
+\partial_{\alpha}h^{\mu\nu}g_{\mu\delta}g_{\nu\beta}\right]
\left[\sum_{n=1}^{z}\tilde{\lambda}_{2,n}n(\tilde{\lambda}_{2,0}h_{\mu\nu}\dot{x}^{\mu}\dot{x}^{\nu})^{n-1}\right]\dot{x}^{\delta}\dot{x}^{\beta}
\]
Finally, putting everything together we have
\[
\dot{x}^{\beta}\nabla_{\beta}\dot{x}^{\alpha}\simeq
\left[
-\partial_{\beta}g^{\alpha\nu}\dot{x}^{\beta}\sum_{n=1}^{z}\tilde{\lambda}_{2,n}n(\tilde{\lambda}_{2,0}h_{\mu\nu}\dot{x}^{\mu}\dot{x}^{\nu})^{n-1}h^{\phantom{-}\gamma}_{\nu}g_{\gamma\delta}\dot{x}^{\delta}\right]+
\]\[
+\sum_{n=1}^{z}\tilde{\lambda}_{2,n}n\left[\left(-g^{\alpha\nu}\partial_{\nu}g_{\mu\beta}h^{\mu\gamma}g_{\delta\gamma}
-\frac{1}{2}g^{\alpha\nu}\partial_{\nu}h^{\mu\gamma}g_{\mu\delta}g_{\gamma\beta}\right)
(\tilde{\lambda}_{2,0}h_{\mu\nu}\dot{x}^{\mu}\dot{x}^{\nu})^{n-1}\dot{x}^{\delta}\dot{x}^{\beta}+\right.
\]\[
+\left.(n-1)(\tilde{\lambda}_{2,0}h_{\mu\nu}\dot{x}^{\mu}\dot{x}^{\nu})^{n-2}(\tilde{\lambda}_{2,0}\dot{h}^{\beta\gamma}g_{\beta\delta}g_{\gamma\rho}\dot{x}^{\rho}\dot{x}^{\delta}+\tilde{\lambda}_{2,0}h^{\beta\gamma}\partial_{\beta}g_{\delta\rho}\dot{x}^{\delta}\dot{x}^{\rho}g_{\gamma\sigma}\dot{x}^{\sigma})h^{\alpha\nu}g_{\nu\phi}\dot{x}^{\phi}+\right.
\]\[
\left.+(\tilde{\lambda}_{2,0}h_{\mu\nu}\dot{x}^{\mu}\dot{x}^{\nu})^{n-1}
(\dot{h}^{\alpha\nu}g_{\nu\delta}\dot{x}^{\delta}+\frac{1}{2}h^{\alpha\nu}\partial_{\nu}g_{\gamma\delta}\dot{x}^{\gamma}\dot{x}^{\delta})\right]+O(\tilde{\lambda}^{2}).
\]
As expected, the geodesic equation, relative to the metric $g_{\mu\nu}$, is recovered as the zero-order approximation in the $\tilde{\lambda}$'s. In this deformed kinematics the equation of motion depends from the values of the $\tilde{\lambda}$'s and hence each particle, having different $\tilde{\lambda}$'s, will follow a slightly different trajectory; this makes possible to verify experimentally if the kinematics is deformed or not. An interesting feature is that the motion will depend also on the mass of the particle.

Similarly for the massless case $\tilde{\lambda}_{2,0}=0$ we obtain
\[
\dot{x}^{\beta}\nabla_{\beta}\dot{x}^{\alpha}\simeq
-\partial_{\beta}g^{\alpha\nu}\dot{x}^{\beta}\sum_{n=1}^{z}\tilde{\lambda}_{2,n}n(h_{\mu\nu}\dot{x}^{\mu}\dot{x}^{\nu})^{n-1}h^{\phantom{-}\gamma}_{\nu}g_{\gamma\delta}\dot{x}^{\delta}+
\]\[
+\sum_{n=1}^{z}\tilde{\lambda}_{2,n}n\left[\left(-g^{\alpha\nu}\partial_{\nu}g_{\mu\beta}h^{\mu\gamma}g_{\delta\gamma}
-\frac{1}{2}g^{\alpha\nu}\partial_{\nu}h^{\mu\gamma}g_{\mu\delta}g_{\gamma\beta}\right)
(h_{\mu\nu}\dot{x}^{\mu}\dot{x}^{\nu})^{n-1}\dot{x}^{\delta}\dot{x}^{\beta}+\right.
\]\[
+\left.(n-1)(h_{\mu\nu}\dot{x}^{\mu}\dot{x}^{\nu})^{n-2}(\dot{h}^{\beta\gamma}g_{\beta\delta}g_{\gamma\rho}\dot{x}^{\rho}\dot{x}^{\delta}+h^{\beta\gamma}\partial_{\beta}g_{\delta\rho}\dot{x}^{\delta}\dot{x}^{\rho}g_{\gamma\sigma}\dot{x}^{\sigma})h^{\alpha\nu}g_{\nu\phi}\dot{x}^{\phi}+\right.
\]\[
\left.+(h_{\mu\nu}\dot{x}^{\mu}\dot{x}^{\nu})^{n-1}
(\dot{h}^{\alpha\nu}g_{\nu\delta}\dot{x}^{\delta}+\frac{1}{2}h^{\alpha\nu}\partial_{\nu}g_{\gamma\delta}\dot{x}^{\gamma}\dot{x}^{\delta})\right]+O(\tilde{\lambda}^{2}).
\]

\section{The Minkowski Case}\label{sec:Minkowski}
As noted in \cite{KehagiasSfetsos:BHFRWGNRG} the action (\ref{RKSaction}) has as solution the Minkowski vacuum with $g=\{-1,1,1,1\}$ and $n_{\alpha}=(-1,0,0,0)$, that is, with $h=\{0,1,1,1\}$. This solution corresponds to a particular choice of coordinates. In general, we also have to consider how to pass from one coordinate frame system to another. In this section we will call the frame system in which the metric is given by $g=\{-1,1,1,1\}$ and $h=\{0,1,1,1\}$ the ``\emph{preferred frame system}'' and we will introduce the notion of ``\emph{inertial frame systems}'' which will have the same Minkowskian metric $g=\{-1,1,1,1\}$ but a different $h^{\mu\nu}$. The group of coordinate transformations which leave invariant the Minkowski metric is the Poincar\'e group, but the ray optical structure is invariant only under rotations and translations\footnote{The linear transformations of coordinates that leave unchanged the metric $g=\{-1,1,1,1\}$ and the vector $n_{\alpha}=(-1,0,0,0)$ are spatial rotations and spacial and time translations:
\begin{center}
$\mathbf{x}'=R\mathbf{x}+x_{0} \quad t'=t+t_{0}.$
\end{center}}. In this context, although the metric remains invariant, we move from the preferred frame system in which $n_{\alpha}=(-1,0,0,0)$ to others in which this vector, and consequently $h^{\mu\nu}$, will be different. The dynamics of particles will be different in these frames allowing, in principle, the possibility to distinguish between any inertial frame system and the preferred inertial frame system. Therefore, in such a theory the preferred frame system plays the role of an absolute frame system.

\subsection{Massive particles in the ``Preferred Frame System''}
Let us start considering the equation of motion in the preferred frame system ($h=\{0,1,1,1\}$); moreover we will consider for simplicity and w.l.o.g.\footnote{The action is invariant under the spacial rotation group.} that the particle is moving along the $x$ direction and that $p_{y}=p_{z}=0$. Then, the equations (\ref{EMpmassive},\ref{EMxmassive}) reduce to
\begin{equation}\label{PFS_x_dot}
\dot{t}=-\frac{p_{t}}{\sqrt{\tilde{\lambda}_{2,0}}}
\quad
\dot{x}=\frac{1}{\sqrt{\tilde{\lambda}_{2,0}}}\left\{1+\sum_{n=1}^{z}\tilde{\lambda}_{2,n}np_{x}^{2(n-1)}\right\}p_{x}
\quad
\dot{y}=0 \qquad \dot{z}=0
\end{equation}
and
\begin{equation}\label{PFS_p_dot}
\dot{p}_{\alpha}=-\frac{\partial H}{\partial x^{\alpha}}=0.
\end{equation}
Note that in this case the conditions $p_{\alpha}=0$ and $\dot{x}^{\alpha}=0$ are consistent; that is, when a particle is at rest the linear momentum is zero, which is not the case in other frame systems.

Integrating with respect to the parameter $\tau$, we obtain the trajectories
\[
t(\tau)=-\frac{p_{t}}{\sqrt{\tilde{\lambda}_{2,0}}}\tau+t_{0}
\quad
x(\tau)=\left[1+\sum_{n=1}^{z}\tilde{\lambda}_{2,n}np_{x}^{2(n-1)}\right]\frac{1}{\sqrt{\tilde{\lambda}_{2,0}}}p_{x}\tau+x_{0}
\quad
\begin{array}{c}
y(\tau)=y_{0} \\ \\ z(\tau)=z_{0}
\end{array}
\]
with the dispersion relation following from (\ref{Hmassive})
\begin{equation}\label{Mdispersionrelation}
-p_{t}^{2}+p_{x}^{2}+\tilde{\lambda}_{2,0}+\sum_{n=1}^{z}\tilde{\lambda}_{2,n}p_{x}^{2n}=0.
\end{equation}
The momentum $p_{x}^{2}$ can be bounded from above or not depending on the values of the $\tilde{\lambda}_{2,n}$'s as is evident from the consistency condition\footnote{
From the dispersion relation we have
\[
-p_{t}^{2}+p_{x}^{2}=-\tilde{\lambda}_{2,0}-\sum_{n=1}^{z}\tilde{\lambda}_{2,n}p_{x}^{2n}
\]
that tells us that there is no condition on the four momentum to be space-, time- or light-like.}
\begin{equation}\label{consistency_condition_p_x}
p_{t}^{2}=p_{x}^{2}+\tilde{\lambda}_{2,0}+\sum_{n=1}^{z}\tilde{\lambda}_{2,n}p_{x}^{2n}\geq0.
\end{equation}
If the preferred frame is the rest frame system of a particle we have $p_{x}=0$ and, from the dispersion relation, $p_{t}=-\sqrt{\tilde{\lambda}_{2,0}}$. In this case $\tau$ can be interpreted as the proper time for this particle, but in general $\tau$ will be just a parameter.

To study how the motion changes moving from one ``inertial frame system'' to another we need to understand the kinematics of particles moving with constant velocity in the preferred frame system, that is, to solve $p_{x}$ as a function of the constant velocity $v=\frac{\dot{x}}{\dot{t}}$. From the equation of motion (\ref{PFS_x_dot}) we have:
\begin{equation}\label{pt=f(px)}
v_{x}=\frac{dx}{dt}=
-\left[1+\sum_{n=1}^{z}\tilde{\lambda}_{2,n}np_{x}^{2(n-1)}\right]\frac{p_{x}}{p_{t}}
\quad\to\quad
p_{t}=
-\left[1+\sum_{n=1}^{z}\tilde{\lambda}_{2,n}np_{x}^{2(n-1)}\right]\frac{p_{x}}{v}.
\end{equation}
Inserting the last relation in the dispersion relation (\ref{Mdispersionrelation}), without any approximation, we have
\[
-p_{x}^{2}\frac{1}{\gamma^{2}v^{2}}+\tilde{\lambda}_{2,0}+\sum_{n=1}^{z}\tilde{\lambda}_{2,n}\left(1-\frac{2n}{v^{2}}\right)p_{x}^{2n}-\left(\sum_{n=1}^{z}\tilde{\lambda}_{2,n}np_{x}^{2(n-1)}\right)^{2}\frac{p_{x}^{2}}{v^{2}}=0
\]
which, approximated to the first order in the $\tilde{\lambda}_{2,n}$'s and for $v<<1$ ($v<<c$), becomes
\[
p_{x}^{2}-\tilde{\lambda}_{2,0}\gamma^{2}v^{2}-\sum_{n=1}^{z}\tilde{\lambda}_{2,n}\gamma^{2}(v^{2}-2n)p_{x}^{2n}\simeq0
\qquad \left(\gamma^{2}\equiv\frac{1}{1-v^{2}}\right).
\]

Without any approximation we have
\begin{equation}\label{minkowski_mass_v}
v^{2}=\left[1+\sum_{n=1}^{z}\tilde{\lambda}_{2,n}np_{x}^{2(n-1)}\right]^{2}\frac{p_{x}^{2}}{\tilde{\lambda}_{2,0}+p_{x}^{2}+\sum_{n=1}^{z}\tilde{\lambda}_{2,n}p_{x}^{2n}};
\end{equation}
Note that the behavior of $v^{2}$ depends from the sign of the $\tilde{\lambda}_{2,n}$'s and, although it must be always positive under the condition (\ref{consistency_condition_p_x}), it may not be a monotonic increasing function of $p_{x}$, depending on the sign of the $\tilde{\lambda}_{2,n}$'s.
In the zero order expansion in the $\tilde{\lambda}_{2,n}$'s (\ref{minkowski_mass_v}) reduces to
\[
v^{2}=\frac{p_{x}^{2}}{\tilde{\lambda}_{2,0}+p_{x}^{2}}
\]
which corresponds to the relativistic relation
\[
p_{x}=\tilde{\lambda}_{2,0}\gamma^{2}v^{2}
\]
reproducing the expected behavior. Now we can use the Newton algorithm to find the roots of a polynomial using as starting point the value $\tilde{\lambda}_{2,0}\gamma^{2}v^{2}$, the other terms of the polynomial being small corrections. After the first step\footnote{
\[
(p^{2})_{0}=\tilde{\lambda}_{2,0}\gamma^{2}v^{2} \quad
y_{0}=-\sum_{n=1}^{z}\tilde{\lambda}_{2,n}(\tilde{\lambda}_{2,0}\gamma^{2}v^{2})^{n}\gamma^{2}[v^{2}-2n]
\]\[
m=\left.\frac{\partial y}{\partial p^{2}}\right|_{(p^{2})_{0}}=
1-\sum_{n=1}^{z}\tilde{\lambda}_{2,n}(\tilde{\lambda}_{2,0}\gamma^{2}v^{2})^{n-1}\gamma^{2}n[v^{2}-2n]
\]
} we have
\[
p_{x}^{2}\simeq
\tilde{\lambda}_{2,0}\gamma^{2}v^{2}
-\frac{-\sum_{n=1}^{z}\tilde{\lambda}_{2,n}(\tilde{\lambda}_{2,0}\gamma^{2}v^{2})^{n}\gamma^{2}[v^{2}-2n]}
{1-\sum_{n=1}^{z}\tilde{\lambda}_{2,n}(\tilde{\lambda}_{2,0}\gamma^{2}v^{2})^{n-1}\gamma^{2}n[v^{2}-2n]}\simeq
\]\[
\tilde{\lambda}_{2,0}\gamma^{2}v^{2}
+\sum_{n=1}^{z}\tilde{\lambda}_{2,n}(\tilde{\lambda}_{2,0}\gamma^{2}v^{2})^{n}\gamma^{2}[v^{2}-2n].
\]
Therefore, $p_{x}$ and $p_{t}$  of a particle moving at a constant velocity $v$ in the preferred frame system are approximated by
\begin{eqnarray}
p_{x} &\simeq&
\left[1+\frac{1}{2}\sum_{n=1}^{z}\tilde{\lambda}_{2,n}(\tilde{\lambda}_{2,0}\gamma^{2}v^{2})^{n-1}\gamma^{2}[v^{2}-2n]\right]\sqrt{\tilde{\lambda}_{2,0}}\gamma v\\
p_{t} &\simeq&
-\left[1-\frac{1}{2\tilde{\lambda}_{2,0}}\sum_{n=1}^{z}\tilde{\lambda}_{2,n}(\tilde{\lambda}_{2,0}\gamma^{2}v^{2})^{n}[2n-1]\right]\sqrt{\tilde{\lambda}_{2,0}}\gamma
\end{eqnarray}
where to evaluate $p_{t}$ we used the relation (\ref{pt=f(px)}). The approximate equations of motion then become
\begin{eqnarray}
t(\tau) &\simeq& \left[1-\frac{1}{2\tilde{\lambda}_{2,0}}\sum_{n=1}^{z}\tilde{\lambda}_{2,n}(\tilde{\lambda}_{2,0}\gamma^{2}v^{2})^{n}[2n-1]\right]\gamma\tau+t_{0}
\\
x(\tau) &\simeq& \left[1-\frac{1}{2\tilde{\lambda}_{2,0}}\sum_{n=1}^{z}\tilde{\lambda}_{2,n}(\tilde{\lambda}_{2,0}\gamma^{2}v^{2})^{n}[2n-1]\right]\gamma v\tau+x_{0}
\end{eqnarray}

In section \ref{SSec:PMGIFS} we will obtain the equations of motion in a generic inertial frame system but we need before to construct operationally a notion of ``inertial frame''; this will be done in the next section.

\subsection{Massless Particles}
The equations of motion for a massless particle ($\tilde{\lambda}_{2,0}=0$) moving along the $x$-axis in the preferred frame system, as derived from the Hamiltonian (\ref{Hmassless}) are
\[
\dot{t}=-p_{t} \qquad
\dot{x}=
\left\{1+\sum_{n=1}^{z}\tilde{\lambda}_{2,n}n(p_{x})^{2(n-1)}\right\}p_{x}
\qquad
\dot{y}=0 \qquad \dot{z}=0.
\]
Following the same procedure as in the massive case, in the massless case we find
\begin{equation}
v_{x}=\frac{dx}{dt}=
-\left[1+\sum_{n=1}^{z}\tilde{\lambda}_{2,n}np_{x}^{2(n-1)}\right]\frac{p_{x}}{p_{t}}
\quad\to\quad
p_{t}=-\left[1+\sum_{n=1}^{z}\tilde{\lambda}_{2,n}np_{x}^{2(n-1)}\right]\frac{p_{x}}{v}.
\end{equation}
Inserting the last relation into the dispersion relation we have
\[
v^{2}=\frac{\left[1+\sum_{n=1}^{z}\tilde{\lambda}_{2,n}np_{x}^{2(n-1)}\right]^{2}}{1+\sum_{n=1}^{z}\tilde{\lambda}_{2,n}p_{x}^{2(n-1)}}\simeq
\left[1+\sum_{n=1}^{z}\tilde{\lambda}_{2,n}(2n-1)p_{x}^{2(n-1)}\right].
\]

It is evident that a massless particle with small energy behaves like in special relativity, that is,
\[
-p_{t}=|p_{x}|
\]
since in this range no $\tilde{\lambda}$'s appear in the relations, and $v^{2}=1$. We can use this property to define ``inertial'' frame systems as the frames in which the massless particles move with constant velocity $v=1$. Moreover, in a ``non-inertial'' frame system the metric will not be anymore Minkowsian and hence the equations of motion will be different from
\begin{equation}\label{EMtest_massless}
\dot{x}=p_{x}
\qquad
\dot{t}=-p_{t}.
\end{equation}

To define an ``inertial'' frame system we start with a frame in which a test massless particle with low energy travels at a constant speed. We define a unit, let us say for length, and we construct the unit of time in such way that, by definition, the velocity of our test particle is $v=1$.

Such a definition of ``inertial'' frame system leads us to consider only frame systems obtained by applying the usual Lorentz transformations\footnote{To measure a constant velocity of $1$ for a massless particle is not enough to fix the units of a frame system because we can always rescale our space-time units by the same constant factor. We can take as more appropriate definition for the choice of coordinates in an other ``inertial frame system'' the unique coordinates and units obtained using Lorentz transformations, once we fixed the units in one of the ``inertial'' frame systems.}. Indeed, only such transformations preserve the relation
\[
\dot{t}^{2}-\dot{x}^{2}=p_{t}^{2}-p_{x}^{2}=0.
\]
This, in particular, means that in such frame systems lengths and time intervals change as in special relativity.

Let us consider for example a boost with velocity $u$ along the $x$-direction. Then the equations of motion change as follows:
\[
\dot{x}'=\gamma(\dot{x}-u\dot{t})=\gamma(p_{x}-up_{t})
\qquad
\dot{t}'=\gamma(\dot{t}-u\dot{x})=\gamma(p_{t}-up_{x})
\]
and hence the equation of motion of our test particle. Considering that in the new frame system we have the same kind of equations of motion as (\ref{EMtest_massless}), we deduce that
\[
p_{x}'=\gamma(p_{x}-up_{t})
\qquad
p_{t}'=\gamma(p_{t}-up_{x})
\]
exactly as in special relativity.

\subsection{Particle Motion in a Generic ``Inertial'' Frame System}\label{SSec:PMGIFS}
Consider a particle in the origin of the rest frame $O'$ and of the moving frame $O$ at $t_{0}=t_{0}'=0$ and that these frames have parallel spatial axes and the moving frame is moving with a velocity $-u$ along the $x$-direction. Consider for the moment $O'$ to be the preferred frame (${h'}^{\alpha\beta}=\{0,1,1,1\}$). The equations of motion in $O'$ for the particle are derived from the optical structure
\[
H=\frac{1}{2\sqrt{\tilde{\lambda}_{2,0}}}\left\{{\eta'}^{\mu\nu}p'_{\mu}p'_{\nu}+\tilde{\lambda}_{2,0}+\sum_{n=1}^{z}\tilde{\lambda}_{2,n}({h'}^{\mu\nu}p'_{\mu}p'_{\nu})^{n}\right\}.
\]
Using the fact that the new coordinates are related to the primed ones by the boost
\[
x=\gamma(x'+ut') \qquad
t=\gamma(t'+ux')
\]
and that the quantities appearing in $H$ are vectors and tensors, and hence transform with the matrix $\frac{\partial {x'}^{\alpha}}{\partial x^{\beta}}$, we have that, in $O$, the metric becomes
\begin{equation}\label{metric-u}
h^{\alpha\beta}=\left(\begin{array}{cccc}
\gamma^{2}u^{2} & \gamma^{2}u & 0 & 0\\
\gamma^{2}u & \gamma^{2} & 0 & 0\\
0 & 0 & 1 & 0\\
0 & 0 & 0 & 1
\end{array}\right) \qquad
\eta^{\alpha\beta}={\eta'}^{\alpha\beta} \qquad
n_{\alpha}=(-\gamma,\gamma u,0,0)
\end{equation}
and $H$ in the new frame system can be written as\footnote{Note that also in this case the approximation $\nabla_{\alpha}h^{\alpha\beta}\simeq0$ holds as an identity leaving the form of the optical structure $H$ unchanged. This is obviously true in the Minkowskian solution for any change of coordinates.}
\[
H=\frac{1}{2\sqrt{\tilde{\lambda}_{2,0}}}\left\{-p_{t}^{2}+p_{x}^{2}+\tilde{\lambda}_{2,0}+\sum_{n=1}^{z}\tilde{\lambda}_{2,n}[\gamma^{2}(up_{t}+p_{x})^{2}]^{n}\right\}
\]
leading to the following equations of motion:
\[
\dot{t}=\frac{1}{\sqrt{\tilde{\lambda}_{2,0}}}\left\{-p_{t}+\sum_{n=1}^{z}\tilde{\lambda}_{2,n}n[\gamma(up_{t}+p_{x})]^{2n-1}\gamma u\right\}
\]\[
\dot{x}=\frac{1}{\sqrt{\tilde{\lambda}_{2,0}}}\left\{p_{x}+\sum_{n=1}^{z}\tilde{\lambda}_{2,n}n[\gamma(up_{t}+p_{x})]^{2n-1}\gamma\right\}
\quad \dot{y}=0 \quad \dot{z}=0
\]
with all the momenta constant and related to the primed momenta, being covectors, by the same transformation rule, that is
\[
p_{t}=\gamma(p'_{t}-up'_{x})=-\gamma\sqrt{\tilde{\lambda}_{2,0}}
\quad
p_{x}=\gamma(p'_{x}-up'_{t})=\gamma\sqrt{\tilde{\lambda}_{2,0}}u.
\]
This means that $up_{t}+p_{x}=p'_{x}=0$ and hence the kinematics is described by the equation of motion
\[
\dot{t}=-\frac{p_{t}}{\sqrt{\tilde{\lambda}_{2,0}}} \quad
\dot{x}=\frac{p_{x}}{\sqrt{\tilde{\lambda}_{2,0}}}
\quad \dot{y}=0 \quad \dot{z}=0
\]
with the usual dispersion relation
\[
-p_{t}^{2}+p_{x}^{2}=0.
\]

Now let us consider the same situation with an $O'$, the rest frame of the particle, which is not the preferred absolute frame system but itself is moving with a constant velocity $-v^{(r)}$ respect to the absolute frame system. Using the fact that in $O'$ the metric takes the same form (\ref{metric-u}) with $v^{(r)}$ instead of $u$, labeling the four-momentum in $O'$ with ${}^{(r)}$, the equations of motion in $O'$ are given by
\begin{equation}\label{EMmoving_rest_frame}
\begin{array}{rcl}
\dot{t}' &=& \frac{1}{\sqrt{\tilde{\lambda}_{2,0}}}\left\{-p^{(r)}_{t}+\sum_{n=1}^{z}\tilde{\lambda}_{2,n}n[\gamma^{(r)}(v^{(r)}p^{(r)}_{t}+p^{(r)}_{x})]^{2n-1}\gamma^{(r)}v^{(r)}\right\}
\\
\dot{x}' &=& \frac{1}{\sqrt{\tilde{\lambda}_{2,0}}}\left\{p^{(r)}_{x}+\sum_{n=1}^{z}\tilde{\lambda}_{2,n}n[\gamma^{(r)}(v^{(r)}p^{(r)}_{t}+p^{(r)}_{x})]^{2n-1}\gamma^{(r)}\right\}
\quad \dot{y}'=0 \quad \dot{z}'=0
\end{array}
\end{equation}
with the dispersion relation
\[
-{p^{(r)}}_{t}^{2}+{p^{(r)}}_{x}^{2}+\tilde{\lambda}_{2,0}+\sum_{n=1}^{z}\tilde{\lambda}_{2,n}[{\gamma^{(r)}}^{2}(v^{(r)}p^{(r)}_{t}+p^{(r)}_{x})^{2}]^{n}=0.
\]
Since the particle is in its rest frame, we have the condition $\dot{x}'=0$; from this condition, using the Newton method with starting point $p^{(r)}_{x}=0$, we obtain
\[
p^{(r)}_{x}\simeq-\frac{\sum_{n=1}^{z}\tilde{\lambda}_{2,n}n({\gamma^{(r)}}^{2}{v^{(r)}}^{2}{p^{(r)}}_{t}^{2})^{n-1}{\gamma^{(r)}}^{2}v^{(r)}p^{(r)}_{t}}{1+\sum_{n=1}^{z}\tilde{\lambda}_{2,n}n(2n-1)({\gamma^{(r)}}^{2}{v^{(r)}}^{2}{p^{(r)}}_{t}^{2})^{n-1}{\gamma^{(r)}}^{2}}\simeq
\]\[
-\sum_{n=1}^{z}\tilde{\lambda}_{2,n}n({\gamma^{(r)}}^{2}{v^{(r)}}^{2}{p^{(r)}}_{t}^{2})^{n-1}{\gamma^{(r)}}^{2}v^{(r)}p^{(r)}_{t}.
\]
We can plug this approximate result into the dispersion relation to find $p_{t}$ using the Newton method with starting point ${p^{(r)}}_{t}^{2}=\tilde{\lambda}_{2,0}$ obtaining
\begin{equation}\label{p^r}
\begin{array}{rcl}
p_{t}^{(r)} &\simeq& -\sqrt{\tilde{\lambda}_{2,0}}\left[1+\frac{1}{2\tilde{\lambda}_{2,0}}\sum_{n=1}^{z}\tilde{\lambda}_{2,n}({\gamma^{(r)}}^{2}{v^{(r)}}^{2}\tilde{\lambda}_{2,0})^{n}\right]
\\
p_{x}^{(r)} &\simeq& -\sqrt{\tilde{\lambda}_{2,0}}\sum_{n=1}^{z}\tilde{\lambda}_{2,n}n({\gamma^{(r)}}^{2}{v^{(r)}}^{2}\tilde{\lambda}_{2,0})^{n-1}{\gamma^{(r)}}^{2}v^{(r)}.
\end{array}
\end{equation}
It is evident then, that only particles at rest in the preferred absolute frame have Lorentz-like dispersion relations and that, in general, the linear momentum is not zero, even if the particle is at rest. Moreover the dispersion relations depend from the velocity $v^{(r)}$, the velocity that the preferred frame system has with respect to the rest frame system of the particle.

Boosting from a non-preferred ``inertial'' rest frame to another will then make the kinematics, through the momenta, dependent on the $\lambda$'s making possible the experimental verification of the existence of such particle parameters.

To evaluate the equation of motion of a particle in a generic ``inertial'' frame system we start with the previous result, that is, we consider given values of $p_{x}^{(r)}$ and $p_{t}^{(r)}$ in the particle rest frame. Noting that in the rest frame we have the condition $\dot{x}^{(r)}=0$ we simply deduce that
\begin{equation}
p_{x}^{(r)}=-\sum_{n=1}^{z}\tilde{\lambda}_{2,n}n[\gamma^{(r)}(v^{(r)}p_{t}^{(r)}+p_{x}^{(r)})]^{2n-1}\gamma^{(r)}
=-\sum_{n=1}^{z}\tilde{\lambda}_{2,n}n[h^{\mu\nu}p_{\mu}p_{\nu}]^{n-1/2}\gamma^{(r)}
\end{equation}
where $v^{(r)}$ is the relative velocity between the ``inertial'' rest frame of the particle and the preferred frame system. Thus the equation of motion in a generic ``inertial'' frame system for a particle moving with a velocity $u$, without any approximation, are
\begin{equation}\label{EMmoving_frame}
\begin{array}{rcl}
\dot{t} &=& \frac{1}{\sqrt{\tilde{\lambda}_{2,0}}}\left\{-p_{t}-\frac{p_{x}^{(r)}}{\gamma^{(r)}}\gamma^{(T)}u^{(T)}\right\}=
\frac{1}{\sqrt{\tilde{\lambda}_{2,0}}}\left\{-p_{t}-p_{x}^{(r)}\gamma_{u}(v^{(r)}+u)\right\}
\\
\dot{x} &=& \frac{1}{\sqrt{\tilde{\lambda}_{2,0}}}\left\{p_{x}-\frac{p_{x}^{(r)}}{\gamma^{(r)}}\gamma^{(T)}\right\}=
\frac{1}{\sqrt{\tilde{\lambda}_{2,0}}}\left\{p_{x}-p_{x}^{(r)}\gamma_{u}(1+uv^{(r)})\right\}
\quad \dot{y}=0 \quad \dot{z}=0
\end{array}
\end{equation}
where
\[
u^{(T)}=\frac{v^{(r)}+u}{1+v^{(r)}u} \quad
\gamma^{(T)}=\gamma^{(r)}\gamma_{u}(1+uv^{(r)}).
\]

In the massless case we can follow the same procedure arriving to the following equations of motion:
\begin{equation}\label{EMmoving_frame_massless}
\begin{array}{rcl}
\dot{t} &=& \left\{-p_{t}-\frac{p_{x}^{(r)}}{\gamma^{(r)}}\gamma^{(T)}u^{(T)}\right\}=
\left\{-p_{t}-p_{x}^{(r)}\gamma_{u}(v^{(r)}+u)\right\}
\\
\dot{x} &=& \left\{p_{x}-\frac{p_{x}^{(r)}}{\gamma^{(r)}}\gamma^{(T)}\right\}=
\left\{p_{x}-p_{x}^{(r)}\gamma_{u}(1+uv^{(r)})\right\}
\quad \dot{y}=0 \quad \dot{z}=0
\end{array}
\end{equation}
The expression for $p_{x}^{(r)}$ and $p_{t}^{(r)}$, however will be different. As in the massive case from the condition $\dot{x}=0$ we obtain
\[
p^{(r)}_{x}\simeq-\frac{\sum_{n=1}^{z}\tilde{\lambda}_{2,n}n({\gamma^{(r)}}^{2}{v^{(r)}}^{2}{p^{(r)}}_{t}^{2})^{n-1}{\gamma^{(r)}}^{2}v^{(r)}p^{(r)}_{t}}{1+\sum_{n=1}^{z}\tilde{\lambda}_{2,n}n(2n-1)({\gamma^{(r)}}^{2}{v^{(r)}}^{2}{p^{(r)}}_{t}^{2})^{n-1}{\gamma^{(r)}}^{2}}\simeq
\]\[
-\sum_{n=1}^{z}\tilde{\lambda}_{2,n}n({\gamma^{(r)}}^{2}{v^{(r)}}^{2}{p^{(r)}}_{t}^{2})^{n-1}{\gamma^{(r)}}^{2}v^{(r)}p^{(r)}_{t}.
\]
which, substituted in the dispersion relation
\[
-{p^{(r)}}_{t}^{2}+{p^{(r)}}_{x}^{2}+\sum_{n=1}^{z}\tilde{\lambda}_{2,n}[{\gamma^{(r)}}^{2}(v^{(r)}p^{(r)}_{t}+p^{(r)}_{x})^{2}]^{n}=0,
\]
yields to the first order in the $\tilde{\lambda}$'s
\[
-{p^{(r)}}_{t}^{2}+\sum_{n=1}^{z}\tilde{\lambda}_{2,n}[{\gamma^{(r)}}v^{(r)}p^{(r)}_{t}]^{2n}=0.
\]
The first solution, $p_{t}^{(r)}=0$, must be rejected because it implies also $p_{x}^{(r)}=0$, that is, the particle is at rest in every ``inertial'' frame. Considering the approximation to the second order the solution $p_{t}^{(r)}=0$ disappear, meaning that must not be considered as a physical solution. The physical solution at the first order corresponds to the roots of the polynomial
\[
\sum_{n=1}^{z}\tilde{\lambda}_{2,n}[{\gamma^{(r)}}v^{(r)}p^{(r)}_{t}]^{2(n-1)}=1.
\]
In this case as in the case in which we want to consider a solution to the second order we cannot employ the Newton method because we cannot guess an initial value for the root. Furthermore, we cannot even consider the $\tilde{\lambda}$'s to be small in general not having any information about massless particles at rest nor having any indication of their existence.

The equations of motion (\ref{EMmoving_frame}) and (\ref{EMmoving_frame_massless}) are the same equations we obtain by boosting the rest frame equation of motion. For example there is no difference, with respect to the prediction of Special Relativity, in the dilation of time
\[
\frac{\Delta t}{\Delta t^{(r)}}=\frac{\dot{t}}{\dot{t}^{(r)}}=
\frac{\left\{-p_{t}-p_{x}^{(r)}\gamma_{u}(v^{(r)}+u)\right\}}{\left\{-p_{t}^{(r)}-p_{x}^{(r)}v^{(r)}\right\}}=
\frac{\left\{-\gamma_{u}(p_{t}^{(r)}-up_{x}^{(r)})-p_{x}^{(r)}\gamma_{u}(v^{(r)}+u)\right\}}{\left\{-p_{t}^{(r)}-p_{x}^{(r)}v^{(r)}\right\}}=
\gamma_{u}
\]
or in the contraction of lengths. But the kinematics of a particle, as is evident from the equations (\ref{EMmoving_frame}) strictly depends from the relative velocity between the ``inertial'' rest frame of the particle and the preferred frame. This will be put in evidence in the next section. Then we will complete the study of the particle motion in section \ref{SSec:LSP} analyzing the general case of luminal and superluminal particles.

\subsection{Scattering}
Suppose we have two identical particles\footnote{Here with identical particles we mean particles with the same $\lambda$'s and the same mass.} in an ``inertial'' frame $O$, with respect to which the preferred frame system $P$ is moving with velocity $u$; the particles $P_{1}$ and $P_{2}$ are moving, respectively, with a velocity $-v$ and a velocity $v$ symmetrically toward the origin $O$. After the collision a unique particle is created. We want to find the dependence of this scattering from the particular ``inertial'' frame system.

The dynamics is described by the Hamiltonian $H=H_{1}+H_{2}$ before the collision and, after, by $H_{T}$. The total conserved time component of the momentum is
\[
(p_{t})_{1}+(p_{t})_{2}=
\gamma_{v}\left\{[(p_{t}^{(r)})_{1}+(p_{t}^{(r)})_{2}]+v[(p_{x}^{(r)})_{1}-(p_{x}^{(r)})_{2}]\right\}=
(p_{t})_{T}
\]
and the conserved spatial component is
\[
(p_{x})_{1}+(p_{x})_{2}=
\gamma_{v}\left\{[(p_{x}^{(r)})_{1}+(p_{x}^{(r)})_{2}]+v[(p_{t}^{(r)})_{1}-(p_{t}^{(r)})_{2}]\right\}=
(p_{x})_{T}.
\]
Now, using the approximate expressions in (\ref{p^r}), we can evaluate the two four-momenta in the rest frame as functions of the relative velocity with the preferred frame:
\begin{equation}
\begin{array}{rcl}
(p_{t}^{(r)})_{1} &\simeq& -\sqrt{\tilde{\lambda}_{2,0}}\left[1+\frac{1}{2\tilde{\lambda}_{2,0}}\sum_{n=1}^{z}\tilde{\lambda}_{2,n}[\gamma_{v}^{2}\gamma_{u}^{2}(u+v)^{2}\tilde{\lambda}_{2,0}]^{n}\right]
\\
(p_{x}^{(r)})_{1} &\simeq& -\sqrt{\tilde{\lambda}_{2,0}}\sum_{n=1}^{z}\tilde{\lambda}_{2,n}n[\gamma_{v}^{2}\gamma_{u}^{2}(u+v)^{2}\tilde{\lambda}_{2,0}]^{n-1}\gamma_{v}^{2}\gamma_{u}^{2}(1+uv)(-u-v)
\\
(p_{t}^{(r)})_{2} &\simeq& -\sqrt{\tilde{\lambda}_{2,0}}\left[1+\frac{1}{2\tilde{\lambda}_{2,0}}\sum_{n=1}^{z}\tilde{\lambda}_{2,n}[\gamma_{v}^{2}\gamma_{u}^{2}(v-u)^{2}\tilde{\lambda}_{2,0}]^{n}\right]
\\
(p_{x}^{(r)})_{2} &\simeq& -\sqrt{\tilde{\lambda}_{2,0}}\sum_{n=1}^{z}\tilde{\lambda}_{2,n}n[\gamma_{v}^{2}\gamma_{u}^{2}(v-u)^{2}\tilde{\lambda}_{2,0}]^{n-1}\gamma_{v}^{2}\gamma_{u}^{2}(1-uv)(v-u).
\end{array}
\end{equation}
where we used the relativistic sum of the velocities
\[
v_{1}^{(P)}=\frac{-v-u}{1+uv} \qquad
v_{2}^{(P)}=\frac{v-u}{1-uv}.
\]
The approximate $(p_{t})_{T}$ is then given by
\[
(p_{t})_{T}=
\gamma_{v}\left\{[(p_{t}^{(r)})_{1}+(p_{t}^{(r)})_{2}]+v[(p_{x}^{(r)})_{1}-(p_{x}^{(r)})_{2}]\right\}\simeq
\]\[
-\sqrt{\tilde{\lambda}_{2,0}}\gamma_{v}\left\{2\left[1
+\frac{1}{2\tilde{\lambda}_{2,0}}\sum_{n=1}^{z}\tilde{\lambda}_{2,n}[\gamma_{v}^{2}\gamma_{u}^{2}\tilde{\lambda}_{2,0}]^{n}
\sum_{k=0}^{n}{2n\choose 2k}u^{2(n-k)}v^{2k}
\right]+\right.
\]\[
\left.-2v\sum_{n=1}^{z}\tilde{\lambda}_{2,n}n[\gamma_{v}^{2}\gamma_{u}^{2}\tilde{\lambda}_{2,0}]^{n-1}\gamma_{v}^{2}\gamma_{u}^{2}
\left(\sum_{k=1}^{n}{2n-1\choose2k-1}u^{2(n-k)}v^{2k-1}+uv\sum_{k=0}^{n-1}{2n-1\choose2k}u^{2(n-k)-1}v^{2k}\right)\right\}=
\]\[
=-2\sqrt{\tilde{\lambda}_{2,0}}\gamma_{v}\left\{1
+\frac{1}{2\tilde{\lambda}_{2,0}}\sum_{n=1}^{z}\tilde{\lambda}_{2,n}[\gamma_{v}^{2}\gamma_{u}^{2}\tilde{\lambda}_{2,0}]^{n}
\sum_{k=0}^{n}\frac{(2n)!}{(2k)!(2n-2k)!}u^{2(n-k)}v^{2k}\left(1-2nv^{2}-2\frac{k}{\gamma_{v}^{2}}\right)
\right\}
\]
while the approximate linear momentum is
\[
(p_{x})_{T}=\gamma_{v}\left\{[(p_{x}^{(r)})_{1}+(p_{x}^{(r)})_{2}]+v[(p_{t}^{(r)})_{1}-(p_{t}^{(r)})_{2}]\right\}\simeq
\]\[
\frac{\gamma_{v}}{\sqrt{\tilde{\lambda}_{2,0}}}\sum_{n=1}^{z}\tilde{\lambda}_{2,n}[\gamma_{v}^{2}\gamma_{u}^{2}\tilde{\lambda}_{2,0}]^{n}
\sum_{k=1}^{2n-1}\frac{(2n)!}{(2k-1)!(2n-2k+1)!}u^{2(n-k)+1}v^{2k-1}\left(\frac{(2k-1)}{v\gamma_{v}^{2}}+(2n-1)v\right)
\]
If we know the final velocity and the $\tilde{\lambda}$'s of the new particle, then we can relate the old $\tilde{\lambda}$'s to the new ones. In general we can expect two possible results: the final particle $P_{T}$ is at rest in $O$ or is moving in $O$. In the case in which the new particle $P_{T}$ is found to be at rest then, noting that the expected four-momentum of the final particle is (\ref{p^r})
\begin{equation}
\begin{array}{rcl}
(p_{t}^{(r)})_{T} &\simeq& -\sqrt{\tilde{\lambda}_{2,0}^{T}}\left[1+\frac{1}{2\tilde{\lambda}_{2,0}^{T}}\sum_{n=1}^{z}\tilde{\lambda}_{2,n}^{T}[\gamma_{u}^{2}u^{2}\tilde{\lambda}_{2,0}^{T}]^{n}\right]
\\
(p_{x}^{(r)})_{T} &\simeq& -\sqrt{\tilde{\lambda}_{2,0}^{T}}\sum_{n=1}^{z}\tilde{\lambda}_{2,n}^{T}n[\gamma_{u}^{2}u^{2}\tilde{\lambda}_{2,0}^{T}]^{n-1}\gamma_{u}^{2}(-u)
\end{array}
\end{equation}
$-u$ being the relative velocity of the particle with respect to the preferred frame, we deduce that, although the mass $2\sqrt{\tilde{\lambda}_{2,0}}\gamma_{v}$ is the same as predicted by special relativity, the other $\lambda$'s come out to be dependent on $u$ as well on $v$. On the other hand, in the case in which the particle $P_{T}$ has a non zero velocity in $O$, we deduce that such a velocity must be first order in the $\tilde{\lambda}$'s; this is so because in the zero order in the $\tilde{\lambda}$'s we expect a particle at rest. Then the velocity of $P_{T}$ and its set of $\tilde{\lambda}$'s will depend on the velocities respect to $O$, the masses, and the $\tilde{\lambda}$'s of the two scattered particles and on the relative velocity of $O$ respect to the preferred frame.

Therefore, in both cases we deduce that physics is different in different ``inertial'' frame systems; that is, the same scattering in two different ``inertial'' frame systems produces two different kinds of particle because the set of $\lambda$'s depends on the relative velocity with the preferred frame system.

\subsection{Luminal and Superluminal Particles}\label{SSec:LSP}
The equations of motion (\ref{EMmoving_frame}) and (\ref{EMmoving_frame_massless}) describe, respectively, massive and massless subluminal particles, being subluminal in every ``inertial'' frame. The motion is determined once we know the four-momentum of the particle in its rest frame, which can be approximately evaluated, knowing all the $\lambda$'s. Also luminal ($|u|=1$) and superluminal particles ($|u|>1$), respectively, are luminal and superluminal in every ``inertial'' frame, but for such particles there does not exist an ``inertial'' rest frame. Therefore, the simplest choice is to write the equations of motion for a generic ``inertial'' frame in terms of the particle four-momentum in the preferred frame.

Then consider the case of a luminal particle in the preferred frame (all other cases can be obtained by appropriate Lorentz transformation). We cannot define the four-momentum in the rest frame for luminal particles, therefore we need to use the equation of motion (\ref{PFS_x_dot}) and the equivalent in the massless case. The condition $u=\frac{\dot{x}}{\dot{t}}=1$, using the equations (\ref{PFS_x_dot}), translates into
\[
\left\{1+\sum_{n=1}^{z}\tilde{\lambda}_{2,n}np_{x}^{2(n-1)}\right\}^{2}p_{x}^{2}=p_{t}^{2}
\]
which, using the dispersion relation (\ref{Mdispersionrelation}), gives
\[
\left[\sum_{n=1}^{z}\tilde{\lambda}_{2,n}np_{x}^{2(n-1)}\right]^{2}p_{x}^{2}
+\sum_{n=1}^{z}\tilde{\lambda}_{2,n}(2n-1)p_{x}^{2n}=\tilde{\lambda}_{2,0}.
\]
The relation above allows us to find $p_{x}$ and then $p_{t}$ in the preferred rest frame. In the massless case it reduces to
\[
\left[\sum_{n=1}^{z}\tilde{\lambda}_{2,n}np_{x}^{2(n-1)}\right]^{2}
+\sum_{n=1}^{z}\tilde{\lambda}_{2,n}(2n-1)p_{x}^{2(n-1)}=0
\]
where we excluded the solution $p_{x}=0$ corresponding to the case of no motion. Knowing $p_{x}$ and then $p_{t}$ in the preferred rest frame means that we can construct the kinematics in any other ``inertial'' frame system. Note that in this case we cannot use the Newton method as in the other cases because we cannot choose an opportune starting point not knowing any expected behavior of the four-momentum nor if the $\tilde{\lambda}$'s are small. Therefore in this case it is necessary to know the $\tilde{\lambda}$'s.

For superluminal particles we can proceed as in the luminal case by considering the four-momentum only in the preferred rest frame because there does not exist any Minkowskian frame in which a superluminal particle is at rest. Then, using the equations of motion (\ref{PFS_x_dot}), the first condition is
\[
\left\{1+\sum_{n=1}^{z}\tilde{\lambda}_{2,n}np_{x}^{2(n-1)}\right\}^{2}p_{x}^{2}=(v^{(P)})^{2}p_{t}^{2}
\]
where $v^{(P)}$ is the velocity of the superluminal particle in the preferred rest frame. The velocity in any other ``inertial'' frame system is obtained with the usual relativistic addition of velocities rule. Using the dispersion relation we have
\[
[1-(v^{(P)})^{2}]p_{t}^{2}+\left[\sum_{n=1}^{z}\tilde{\lambda}_{2,n}np_{x}^{2(n-1)}\right]^{2}p_{x}^{2}
+\sum_{n=1}^{z}\tilde{\lambda}_{2,n}(2n-1)p_{x}^{2n}=\tilde{\lambda}_{2,0}.
\]
We can again find the four-momentum in the preferred rest frame, this time as a function of $v^{(P)}$, allowing us to know the kinematics in any ``inertial'' frame system. In this case also it is necessary to know the $\lambda$'s because we do not have any known behavior of such particles to use as starting point in the Newton method.

We have seen that the deformed kinematics considered here allows in general the presence of superluminal particle and massive luminal particle, depending by the values of the the $\tilde{\lambda}$'s. Indeed from (\ref{minkowski_mass_v}) we have that the velocity can be bounded or unbounded depending from the values of the $\tilde{\lambda}$'s. Moreover, if we have a particle with $\tilde{\lambda}$'s such that the velocity is unbounded, we can accelerate such a particle from rest to superluminal velocities with a constant force.  Consider for example a particle with all the $\tilde{\lambda}$'s positive subject to a constant force $F$ (such effect can be obtained adding the term $-Fx$ to the Hamiltonian changing the equation for $\dot{p}_{x}$ in $\dot{p}_{x}=F$). Then
\[
p_{x}=F\tau=-\sqrt{\tilde{\lambda}_{2,0}}\frac{F}{p_{t}}t
\]
that, using the dispersion relation (\ref{Mdispersionrelation}), gives
\[
p_{x}^{2}[p_{x}^{2}+\tilde{\lambda}_{2,0}+\sum_{n=1}^{z}\tilde{\lambda}_{2,n}p_{x}^{2n}]
+\tilde{\lambda}_{2,0}F^{2}t=0.
\]
Because all the $\tilde{\lambda}$'s are positive the related polynomial is positive and symmetric around $p_{x}=0$ that corresponds to its minimum. For $t>0$ the polynomial develops two symmetric roots which move away from the origin. Therefore in this case $p_{x}$ is a monotonically increasing function of the time. As a result we have that (\ref{minkowski_mass_v}) diverges, therefore we can accelerate a particle with positive $\tilde{\lambda}$'s to any velocity with a constant force.

\subsection{Non-covariant vs Covariant Ho\v{r}ava Theory}
In this section we want to point out the differences between the non-covariant form of the modified Ho\v{r}ava-Lifshitz action (\ref{KSaction}) and its covariant generalization (\ref{RKSaction}). The main difference is that the action (\ref{KSaction}) is written in terms of the fields $h_{ij}$, $N$ , $N^{i}$ while its covariant generalization is in terms of $g_{\alpha\beta}$ and $n_{\alpha}$; this implies that, as described in section \ref{sec:intro}, the covariant action reduces to (\ref{KSaction}) only if it is written in a frame in which the metric $g_{\alpha\beta}$ can be decomposed in terms of $h_{ij}$, $N$ , $N^{i}$.

In the case of the Minkowsky metric such frames are obtained by rotations and translations of the preferred frame system. Indeed, the metric (\ref{metric-u}), obtained by boosting the preferred frame, is not writable in terms of ADM components. Therefore, the action (\ref{RKSaction}) is a generalization of the non-covariant action (\ref{KSaction}) because it includes solutions which are absent in the non-covariant form.

Let $g_{\alpha\beta}$ and $n_{\alpha}$ be a solution of the Kehagias-Sfetsos action (\ref{KSaction}); then $g_{\alpha\beta}$ and $n_{\alpha}$ can be decomposed in terms of its ADM components and it is a solution also for covariant action (\ref{RKSaction}). If we consider a generic change of coordinates $x^{\mu}(x^{\alpha})$, then
\[
g_{\alpha\beta}'=
\frac{\partial x^{\mu}}{\partial x^{'\alpha}}\frac{\partial x^{\nu}}{\partial x^{'\beta}}
g_{\mu\nu}
\]
is another solution for (\ref{RKSaction}) but not for (\ref{KSaction}). If the covariant action (\ref{RKSaction}) is equivalent to the Kehagias-Sfetsos action (\ref{KSaction}), then, after the change of coordinates $x^{\mu}(x^{\alpha})$, the new form assumed by (\ref{KSaction}) can be rewritten in terms $g_{\alpha\beta}'$ and will have the same form as (\ref{RKSaction}).

Our results are independent of the formulation chosen for the Kehagias-Sfetsos action (\ref{KSaction}) because the solutions we considered are solutions of both theories. Moreover the ray optical structure was constructed using these solutions as a background metric. In terms of the optical structure we have that the equation of motion $H[g_{\alpha\beta}(x),n_{\alpha}(x)]=0$, under the coordinate transformation $x^{\mu}(x^{\alpha})$, becomes
\[
H'[g_{\alpha\beta}(x'),n_{\alpha}(x')]=0 \quad\longrightarrow\quad H[g'_{\alpha\beta}(x'),n'_{\alpha}(x')]=0
\]
remaining unchanged, the action (\ref{RKSaction}) being covariant. On the other hand, if we rewrite the same Hamiltonian in terms of its ADM components, that is, $H[h(x),N^{i}(x),$ $N(x)]=0$, the same change of coordinates will change the equation of motion as
\[
H'[h(x'),N^{i}(x'),N(x')]=0.
\]
Such equation is equivalent to $H'[g_{\alpha\beta}(x'),n_{\alpha}(x')]=0$, being just its expansion in terms of the ADM components of $g_{\alpha\beta}$ but cannot be rewritten as 
\[
H[h'(x),N^{'i}(x),N'(x)]=0
\]
 because the transformed metric $g'_{\alpha\beta}$ cannot be decomposed in ADM components.

Therefore, independently from the equivalence of the covariant and the non-covariant form of the modified Ho\v{r}ava action, the optical structure is more conveniently written in a covariant form because the derived equations of motion take a more compact form. The only difference is that, if the non-covariant and the covariant theory are not equivalent, then $g'_{\alpha\beta}$ cannot be interpreted as a metric tensor in the IR limit of Ho\v{r}ava gravity, as it cannot be decomposed in ADM components.

\section{Spherical Symmetric Case}\label{sec:Spherical}
As the original Ho\v{r}ava-Lifshitz theory, the Kehagias-Sfetsos action  (\ref{RKSaction}) possesses a spherical solution \cite{KehagiasSfetsos:BHFRWGNRG}. The solution is based on the following ansatz
\begin{equation}\label{spherical_metric}
g^{\mu\nu}=\left(\begin{array}{cccc}
-\frac{1}{N^{2}} & 0 & 0 & 0\\
0 & f & 0 & 0\\
0 & 0 & \frac{1}{r^{2}} & 0\\
0 & 0 & 0 & \frac{1}{r^{2}\sin^{2}\theta}
\end{array}\right)
\quad
h^{\mu\nu}=\left(\begin{array}{cccc}
0 & 0 & 0 & 0\\
0 & f & 0 & 0\\
0 & 0 & \frac{1}{r^{2}} & 0\\
0 & 0 & 0 & \frac{1}{r^{2}\sin^{2}\theta}
\end{array}\right)
\end{equation}
\[
n^{\alpha}=(-\frac{1}{N},0,0,0)
\]
and the functions $N$ and $f$, for the case $\lambda=1$, are found to be
\begin{equation}\label{f(r)}
N^{2}=f=1+\omega r^{2}-\sqrt{r(\omega^{2}r^{3}+4\omega M)}.
\end{equation}
This solution in the IR limit ($\omega\to\infty$) reproduces the Schwarzschild solution.

In this case $\nabla_{\alpha}h^{\alpha\beta}\simeq0$ does not hold as an identity, its only nonzero component being
\[
\nabla_{\alpha}h^{\alpha r}=\frac{1}{2}\frac{f}{N^{2}}\frac{dN^{2}}{dr}.
\]
Then we can consider the approximation valid only the region in which $\nabla_{\alpha}h^{\alpha\beta}$ is small compared to the wavenumber $1/\lambda$ of the ray. For the solution (\ref{f(r)}) we have
\[
\nabla_{\alpha}h^{\alpha r}=\frac{1}{2}\frac{df}{dr}=
2\omega r-2\omega r\frac{1+\frac{M}{\omega r^{3}}}{\sqrt{1+\frac{4M}{\omega r^{3}}}}\simeq\frac{2M}{r^{2}},
\]
where we approximated to the first order in $1/r^{3}$ in the last step. We can then say that the approximation $\nabla_{\alpha}h^{\alpha\beta}\simeq0$ is valid for $r>>\sqrt{2M\lambda}$.

The metric is a function only of the coordinates $r$ and $\theta$ and diagonal, therefore
\[
\dot{p}_{t}=0 \qquad
\dot{p}_{\phi}=0.
\]
As in the relativistic case, we can define\footnote{Given a vector field $K^{\alpha}$, the product $K_{\alpha}\dot{x}^{\alpha}$ is conserved along a path $x^{\alpha}(\tau)$ if There is no notion of Killing vectors
\[
\frac{d}{d\tau}(K_{\alpha}\dot{x}^{\alpha})=
\dot{x}^{\beta}\nabla_{\beta}(K_{\alpha}\dot{x}^{\alpha})=
\dot{x}^{\beta}\dot{x}^{\alpha}\nabla_{\beta}K_{\alpha}+\dot{x}^{\beta}K_{\alpha}\nabla_{\beta}\dot{x}^{\alpha}=0.
\]
Being $\dot{x}^{\beta}\nabla_{\beta}\dot{x}^{\alpha}\neq0$, we deduce that there is no notion of Killing vectors.}
\[
p_{t}=-E \qquad p_{\phi}=L.
\]
Moreover
\begin{equation}\label{spherical_dot_t}
\dot{t}=
\frac{1}{N^{2}}\frac{1}{\sqrt{\tilde{\lambda}_{2,0}}}E
\end{equation}
and
\[
\dot{\theta}=
\frac{1}{\sqrt{\tilde{\lambda}_{2,0}}}\left[1+\sum_{n=1}^{z}\tilde{\lambda}_{2,n}n(h^{\mu\nu}p_{\mu}p_{\nu})^{n-1}\right]\frac{1}{r^{2}}p_{\theta}
\]\[
\dot{p}_{\theta}=
-\frac{1}{2\sqrt{\tilde{\lambda}_{2,0}}}\left[1+\sum_{n=1}^{z}\tilde{\lambda}_{2,n}n(h^{\rho\lambda}p_{\rho}p_{\lambda})^{n-1}\right]
\left(\frac{-2\cos\theta}{r^{2}\sin^{3}\theta}\right)L^{2}
\]
which have as solution
\[
\theta=\frac{\pi}{2} \qquad p_{\theta}=0.
\]
Therefore, in what follows we will concentrate to the case of equatorial motion. Moreover,
\begin{equation}\label{spherical_dot_p_r}
\dot{p}_{r}=
\frac{1}{2\sqrt{\tilde{\lambda}_{2,0}}}\partial_{r}\left(\frac{1}{N^{2}}\right)E^{2}
-\frac{1}{2\sqrt{\tilde{\lambda}_{2,0}}}\left[1
+\sum_{n=1}^{z}\tilde{\lambda}_{2,n}n(h^{\rho\lambda}p_{\rho}p_{\lambda})^{n-1}\right]\partial_{r}h^{\mu\nu}p_{\mu}p_{\nu}
\end{equation}

\begin{equation}\label{spherical_dot_r}
\dot{r}=
\frac{1}{\sqrt{\tilde{\lambda}_{2,0}}}\left[1
+\sum_{n=1}^{z}\tilde{\lambda}_{2,n}n(h^{\mu\nu}p_{\mu}p_{\nu})^{n-1}\right]f(r)p_{r}
\end{equation}

\begin{equation}\label{spherical_dot_phi}
\dot{\phi}=
\frac{1}{\sqrt{\tilde{\lambda}_{2,0}}}\left[
1+\sum_{n=1}^{z}\tilde{\lambda}_{2,n}n(h^{\mu\nu}p_{\mu}p_{\nu})^{n-1}\right]\frac{L}{r^{2}}.
\end{equation}

From the dispersion relation, on the other hand, we have
\[
0=-\frac{E^{2}}{N^{2}}+\tilde{\lambda}_{2,0}+\left[1+\sum_{n=1}^{z}\tilde{\lambda}_{2,n}\left(f(r)p_{r}^{2}+\frac{L^{2}}{r^{2}}\right)^{n-1}\right]\left[f(r)p_{r}^{2}+\frac{L^{2}}{r^{2}}\right].
\]
To find an approximate result we use again Newton's method. Setting
\[
x=f(r)p_{r}^{2}+\frac{L^{2}}{r^{2}}
\]
we choose as starting point
\[
x_{0}=\frac{E^{2}}{N^{2}}-\tilde{\lambda}_{2,0}
\]
which corresponds to the zero order in the $\lambda$'s. Then we have
\[
y_{0}=\sum_{n=1}^{z}\tilde{\lambda}_{2,n}\left(\frac{E^{2}}{N^{2}}-\tilde{\lambda}_{2,0}\right)^{n}
\quad
m_{0}=\left.\frac{dy}{dx}\right|_{x_{0}}=
1+\sum_{n=1}^{z}\tilde{\lambda}_{2,n}n\left(\frac{E^{2}}{N^{2}}-\tilde{\lambda}_{2,0}\right)^{n-1}.
\]
Therefore,
\[
p_{r}^{2}\simeq
\frac{1}{f(r)}\left\{\frac{E^{2}}{N^{2}}-\tilde{\lambda}_{2,0}-\frac{L^{2}}{r^{2}}-\frac{\sum_{n=1}^{z}\tilde{\lambda}_{2,n}\left(\frac{E^{2}}{N^{2}}-\tilde{\lambda}_{2,0}\right)^{n}}{1+\sum_{n=1}^{z}\tilde{\lambda}_{2,n}n\left(\frac{E^{2}}{N^{2}}-\tilde{\lambda}_{2,0}\right)^{n-1}}\right\}\simeq
\]\[
\frac{1}{f(r)}\left\{\frac{E^{2}}{N^{2}}-\tilde{\lambda}_{2,0}-\frac{L^{2}}{r^{2}}-\sum_{n=1}^{z}\tilde{\lambda}_{2,n}\left(\frac{E^{2}}{N^{2}}-\tilde{\lambda}_{2,0}\right)^{n}\right\}=
\]\[
\frac{1}{f(r)}\left[\frac{E^{2}}{N^{2}}-\tilde{\lambda}_{2,0}-\frac{L^{2}}{r^{2}}\right]\left\{1-\sum_{n=1}^{z}\tilde{\lambda}_{2,n}\frac{\left(\frac{E^{2}}{N^{2}}-\tilde{\lambda}_{2,0}\right)^{n}}{\left[\frac{E^{2}}{N^{2}}-\tilde{\lambda}_{2,0}-\frac{L^{2}}{r^{2}}\right]}\right\}
\]
which yields
\begin{equation}\label{spherical_p_r_approx}
p_{r}\simeq
\frac{1}{\sqrt{f(r)}}\sqrt{\frac{E^{2}}{N^{2}}-\tilde{\lambda}_{2,0}-\frac{L^{2}}{r^{2}}}\left\{1-\frac{1}{2}\sum_{n=1}^{z}\tilde{\lambda}_{2,n}\frac{\left(\frac{E^{2}}{N^{2}}-\tilde{\lambda}_{2,0}\right)^{n}}{\left[\frac{E^{2}}{N^{2}}-\tilde{\lambda}_{2,0}-\frac{L^{2}}{r^{2}}\right]}\right\}
\end{equation}
Using the expressions (\ref{spherical_dot_r}), (\ref{spherical_dot_phi}) and (\ref{spherical_p_r_approx}) we have the approximate orbit equation
\[
\frac{\dot{r}}{\dot{\phi}}=\frac{dr}{d\phi}=
\frac{r^{2}f(r)p_{r}}{L^{2}}.
\]
Orbits as well as other classical gravitational tests were studied in the context of the Kehagias-Sfetsos modification of the Ho\v{r}ava-Lifshitz gravity \cite{Konoplya:TCHLG,HarkoKivacsLobo:SSTHLG,IorioRuggiero:HLGSSOM} showing the presence of corrections with respect to the results of General Relativity. In these articles the kinematic is assumed to be described by the Hamiltonian
\[
H=\frac{1}{2\sqrt{\tilde{\lambda}_{2,0}}}g_{\mu\nu}p^{\mu}p^{\nu}
\]
while we verify this assumption. Our results, being obtained as the optical limit of a scalar field theory, show that the kinematic is indeed described by that Hamiltonian in the zero order in the $\tilde{\lambda}$'s but they receive $\tilde{\lambda}$ dependent correctionS.

In the more general framework allowed by the Ho\v{r}ava-Lifshitz gravity, that is, for non-zero $\tilde{\lambda}$'s we can consider two simple cases which can be used to verify the existence of such parameters: circular and radial orbits.

To have a circular orbit we must have $\dot{r}=0$ and therefore, from (\ref{spherical_dot_r}), $p_{r}=0$. Using the equation (\ref{spherical_dot_p_r}) we have the condition
\[
0=
\partial_{r}\left(\frac{E^{2}}{N^{2}}\right)
-\left[1+\sum_{n=1}^{z}\tilde{\lambda}_{2,n}n\left(\frac{L^{2}}{r^{2}}\right)^{n-1}\right]\left(-2\frac{L^{2}}{r^{3}}\right)=
\partial_{r}\left[\frac{E^{2}}{N^{2}}-\frac{L^{2}}{r^{2}}
-\sum_{n=1}^{z}\tilde{\lambda}_{2,n}\left(\frac{L^{2}}{r^{2}}\right)^{n}\right]
\]
which is satisfied if
\[
\frac{E^{2}}{N^{2}}-\frac{L^{2}}{r^{2}}
-\sum_{n=1}^{z}\tilde{\lambda}_{2,n}\left(\frac{L^{2}}{r^{2}}\right)^{n}=K
\]
where $K$ is a constant. From the dispersion relation we simply deduce $K=\tilde{\lambda}_{2,0}$. From the previous expression then we obtain the energy associated with a circular motion
\[
\frac{E^{2}}{N^{2}}=\tilde{\lambda}_{2,0}+\frac{L^{2}}{r^{2}}+\sum_{n=1}^{z}\tilde{\lambda}_{2,n}\left(\frac{L^{2}}{r^{2}}\right)^{n}.
\]
From the expression (\ref{spherical_dot_r}) and the approximate relation for $p_{r}^{2}$ we obtain:
\[
\dot{r}^{2}\simeq
\frac{f(r)}{\tilde{\lambda}_{2,0}}
\left\{\left[\frac{E^{2}}{N^{2}}-\tilde{\lambda}_{2,0}-\frac{L^{2}}{r^{2}}\right]
-\sum_{n=1}^{z}\tilde{\lambda}_{2,n}\left(\frac{E^{2}}{N^{2}}-\tilde{\lambda}_{2,0}\right)^{n-1}\left[(1-2n)\left(\frac{E^{2}}{N^{2}}-\tilde{\lambda}_{2,0}\right)+2n\frac{L^{2}}{r^{2}}\right]\right\}.
\]
We can interpret the right hand side of this equation as a potential $V(r)$. From the condition $\frac{dV}{dr}=0$ for circular orbits we have
\[
-\frac{E^{2}}{N^{3}}N'+\frac{L^{2}}{r^{3}}
+\sum_{n=1}^{z}\tilde{\lambda}_{2,n}\left(\frac{L^{2}}{r^{2}}\right)^{n-1}
\left[-\frac{E^{2}}{N^{3}}N'(6n-2)-\frac{L^{2}}{r^{3}}2n\right]=0.
\]
So it is evident that deviations for the radius of a circular orbits from the zero-order solution $\frac{E^{2}}{N^{3}}N'=\frac{L^{2}}{r^{3}}$ will increase for increasing angular momentum.

Another simple consequence of this deformed kinematics is that two particles with the same mass and different $\tilde{\lambda}$'s will fall radially toward the center with two different velocities.

From the equation (\ref{spherical_dot_phi}) it is simple to see that a radial solution  corresponds to the case $L=0$. In this case we have the equations of motion become
\[
\dot{p}_{r}=
\frac{1}{2\sqrt{\tilde{\lambda}_{2,0}}}\partial_{r}\left(\frac{1}{N^{2}}\right)E^{2}
-\frac{1}{2\sqrt{\tilde{\lambda}_{2,0}}}\left[1
+\sum_{n=1}^{z}\tilde{\lambda}_{2,n}n[f(r)p_{r}^{2}]^{n-1}\right]\partial_{r}f(r)p_{r}^{2}
\]\[
\dot{r}=
\frac{1}{\sqrt{\tilde{\lambda}_{2,0}}}\left[1
+\sum_{n=1}^{z}\tilde{\lambda}_{2,n}n[f(r)p_{r}^{2}]^{n-1}\right]f(r)p_{r}
\]
and the dispersion relation reduces to
\[
0=-\frac{E^{2}}{N^{2}}+\tilde{\lambda}_{2,0}+\left[1+\sum_{n=1}^{z}\tilde{\lambda}_{2,n}[f(r)p_{r}^{2}]^{n-1}\right]f(r)p_{r}^{2}.
\]
Using the approximate relation (\ref{spherical_p_r_approx}) for $p_{r}$ we have
\[
p_{r}\simeq
\frac{1}{\sqrt{f(r)}}\sqrt{\frac{E^{2}}{N^{2}}-\tilde{\lambda}_{2,0}}\left\{1-\frac{1}{2}\sum_{n=1}^{z}\tilde{\lambda}_{2,n}\left(\frac{E^{2}}{N^{2}}-\tilde{\lambda}_{2,0}\right)^{n-1}\right\}
\]
from which we have
\[
\dot{r}\simeq
\sqrt{\frac{f(r)}{\tilde{\lambda}_{2,0}}}\sqrt{\frac{E^{2}}{N^{2}}-\tilde{\lambda}_{2,0}}\left\{1-\frac{1}{2}\sum_{n=1}^{z}\tilde{\lambda}_{2,n}\left(\frac{E^{2}}{N^{2}}-\tilde{\lambda}_{2,0}\right)^{n-1}\right\}\left[1
+\sum_{n=1}^{z}\tilde{\lambda}_{2,n}n\left(\frac{E^{2}}{N^{2}}-\tilde{\lambda}_{2,0}\right)^{n-1}\right]\simeq
\]\[
\sqrt{\frac{f(r)}{\tilde{\lambda}_{2,0}}}\sqrt{\frac{E^{2}}{N^{2}}-\tilde{\lambda}_{2,0}}
\left[1+\sum_{n=1}^{z}\tilde{\lambda}_{2,n}\left(n-\frac{1}{2}\right)\left(\frac{E^{2}}{N^{2}}-\tilde{\lambda}_{2,0}\right)^{n-1}\right]
\]
Using the equation (\ref{spherical_dot_t}) describing the coordinate time, we have
\[
\frac{dr}{dt}\simeq
\sqrt{f(r)}\frac{N^{2}}{E}\sqrt{\frac{E^{2}}{N^{2}}-\tilde{\lambda}_{2,0}}
\left[1+\sum_{n=1}^{z}\tilde{\lambda}_{2,n}\left(n-\frac{1}{2}\right)\left(\frac{E^{2}}{N^{2}}-\tilde{\lambda}_{2,0}\right)^{n-1}\right].
\]

In general the solution is not expected to be very different from the zero-order case although there are two simple possible measures, the radius of circular orbits and the falling time of a radial orbit, which can show differences. The metric (\ref{spherical_metric}) reduces to the Schwarzschild metric if the mass of the spherical body at the center can be considered small with respect to the distance at which we consider the metric; therefore we can expect to measure variations due to the deformed kinematics only in this approximation where there are negligible deformations in the kinematics due to the Ho\v{r}ava spherically symmetric solution.

\section{Conclusions}
The action (\ref{RSMatterJ=2}) is a deformation of the Klein-Gordon action in the sense that it introduces new interacting terms small in the IR behavior. Such a deformation has nothing to do with the deformation of the Hilbert-Einstein theory used to construct the Ho\v{r}ava-Lifshitz gravity, but it is allowed by the breaking of the Lorentz symmetry of space-time. This means that a scalar may still be described by the usual Klien-Gordon action without any consequence for the whole theory.

The full Ho\v{r}ava-Lifshitz gravity is invariant under local rotations and translations but this symmetry is not enough to construct the kinematics without having any prescription in the addition of the velocities; this is so because there is no mixing between space and time. Then we studied the modification in the kinematics under the action of the Poincar\'e group, mainly for two reasons: the first is that Special Relativity and General Relativity are well verified theories and are based on the Lorentz group; the second reason is related to the first but it is more practical: we used low-energy light, which appears to have the same property that light has in Special Relativity, as test particles to define operationally ``inertial'' frame systems.

We used the action (\ref{RSMatterJ=2}) to analyze the deformed kinematics of a particle, the particle being considered as the optical limit of a scalar field. The deformed equation of motion can be experimentally verified; the results of such experiments may suggest the existence of new parameters describing a particle, the $\tilde{\lambda}'s$, or may give, at least, an upper limit for them.

In the general case we observed that free falling point-particles do not follow geodesics, their trajectories being a small deformation of the geodesic equation. It is also important to observe that such a deformation depends on the values of the $\tilde{\lambda}$'s and even on the mass of the particle. This allows us to verify if a deformed or the usual kinematics is fulfilled even if we use two probes of the same kind but with different masses. In particular we evaluated first order corrections to the orbits in a Schwarzschild metric.

In the Minkowskian case the equations of motion of a free particle still give a straight trajectory, thus it is in studying the dynamics that it is possible to highlight the differences with the usual relativistic dynamics.

We observed that in this modified kinematics there may exist subluminal massless particles, luminal massive particles and superluminal particles; moreover it is possible to accelerate a subluminal particle through the action of a constant force to make the particle to reach superluminal velocities in a finite time, if the particle satisfies appropriate conditions for the $\tilde{\lambda}$'s. It is also evident that the $\tilde{\lambda}$'s of a particle are strictly dependent on the ``inertial'' frame system in which the particle is at rest. Indeed, the result of a scattering will in general depend on the relative velocity between the ``inertial'' frame system in which the resulting particle is at rest and the preferred frame system.

All the observations we already have about the motion of particles in Special Relativity and in General Relativity suggest that the $\tilde{\lambda}$'s are very small parameters or simply that no known matter has nonzero $\tilde{\lambda}$'s.

\vskip 0.2in
\noindent
{\it{Note added}:} 
After completing this manuscript, papers \cite{Suyama:NMHLG,RomeroCuestaGarciaVergara:CAM} appeared in the archives where the kinematics of a matter scalar field in the Ho\v{r}ava-Lifshitz gravity is also studied.

%\vskip 0.2in
%\noindent
%\emph{\underline {Acknowledgements}:} 
%I wish to thank my advisor A.~P.~Polychronakos for all the comments and useful discussions.

\appendix

\section{Variation of the Matter Action Respect to $\phi^{\star}$}\label{sec:appendix}
In the action (\ref{RSMatterJ=2}) the deviation from the Klein-Gordon action is described by the term
\[
-\sum_{n=1}^{z}\sum_{k=0}^{n}(-1)^{n}\frac{\lambda_{2,n,k}}{M^{2(n-1)}}\Delta^{n-k}\phi\Delta^{k}\phi^{*}
\]
where we set $\lambda_{2,1,0}\equiv\lambda_{2,1,1}=\lambda_{2,1}-\frac{1}{2}$. Having in mind to do a variation with respect to $\phi^{\star}$ we have that
\[
\Delta^{n-k}\phi\Delta^{k}\phi^{*}=
\Delta^{n-k}\phi(h^{\mu\nu}\nabla_{\mu}\nabla_{\nu})\Delta^{k-1}\phi^{*}=
\]\[
=\nabla_{\mu}\left[\Delta^{n-k}\phi h^{\mu\nu}\nabla_{\nu}\Delta^{k-1}\phi^{*}\right]
-\nabla_{\mu}\left[\Delta^{n-k}\phi h^{\mu\nu}\right]
\nabla_{\nu}\Delta^{k-1}\phi^{*}=
\]\[
=\nabla_{\mu}\left[\Delta^{n-k}\phi h^{\mu\nu}\nabla_{\nu}\Delta^{k-1}\phi^{*}\right]
-\nabla_{\nu}\left[\nabla_{\mu}\left(\Delta^{n-k}\phi h^{\mu\nu}\right)\Delta^{k-1}\phi^{*}\right]
+\nabla_{\nu}\nabla_{\mu}\left[\Delta^{n-k}\phi h^{\mu\nu}\right]
\Delta^{k-1}\phi^{*}.
\]
Defining the operator $\overrightarrow{\Delta}$ such that its action is given by $\overrightarrow{\Delta}\phi=\nabla_{\alpha}\nabla_{\beta}(h^{\alpha\beta}\phi)$ and considering that all the total covariant derivatives are boundary terms in the action, we conclude that
\[
\Delta^{n-k}\phi\Delta^{k}\phi^{*}=
\overrightarrow{\Delta}\Delta^{n-k}\phi\Delta^{k-1}\phi^{*}.
\]
Therefore, varying with respect to $\phi^{*}$ we have
\[
\delta_{\phi^{*}}\left[\sum_{k=0}^{n}\lambda_{2,n,k}\Delta^{n-k}\phi\Delta^{k}\phi^{*}\right]=
\sum_{k=0}^{n}\lambda_{2,n,k}\overrightarrow{\Delta}^{k}\Delta^{n-k}\phi\delta\phi^{*}+\textrm{total derivativs}.
\]
Using the condition (\ref{approx_metric}), which translates in $\overrightarrow{\Delta}\psi\simeq\Delta\psi$, we conclude that
\[
\delta_{\phi^{*}}\left[\sum_{k=0}^{n}\lambda_{2,n,k}\Delta^{n-k}\phi\Delta^{k}\phi^{*}\right]=
\delta_{\phi^{*}}\left[\sum_{k=0}^{n}\lambda_{2,n,k}\Delta^{n}\phi\phi^{*}\right]=
\lambda_{2,n}\Delta^{n}\phi\delta\phi^{*}.
\]
where we defined $\lambda_{2,n}=\sum_{k=0}^{n}\lambda_{2,n,k}$.

\end{document}